\documentclass[pre,twocolumn,noshowpacs,preprintnumbers,amsmath,amssymb]{revtex4}
\usepackage[dvipdfmx]{graphicx}
\usepackage{color}
\usepackage{mathrsfs}
\usepackage{dcolumn}
\usepackage{bm}
\usepackage{times}
\usepackage{ulem}

\newcommand{\ra}{\rangle}
\newcommand{\la}{\langle}
\newcommand{\mrss}{\mathrm{ss}}
\newcommand{\tilh}{\tilde{h}}
\newcommand{\mrwin}{\mathrm{win}}
\newcommand{\eps}{\epsilon}

\newcommand{\twin}{t_{\mathrm{win}}}
\newcommand{\Dzdiff}{\Delta z_{\mathrm{diff}}}

\begin{document}
\title{Ubiquitous power law scaling in nonlinear self-excited Hawkes processes}

\author{Kiyoshi Kanazawa}
\email{kiyoshi@sk.tsukuba.ac.jp}
\affiliation	{
				Faculty of Engineering, Information and Systems, University of Tsukuba, Tennodai, Tsukuba, Ibaraki 305-8573, Japan\\
				JST, PRESTO, 4-1-8 Honcho, Kawaguchi, Saitama 332-0012, Japan
		}	
\author{Didier Sornette}
\email{dsornette@ethz.ch}
\affiliation	{
				ETH Zurich, Department of Management, Technology, and Economics, Zurich 8092, Switzerland\\
				Institute of Risk Analysis, Prediction, and Management (Risks-X), Academy for Advanced Interdisciplinary Studies, 
				Southern University of Science and Technology (SUSTech), Shenzhen 518055, China
		}
\date{\today}

\begin{abstract}
	The origin(s) of the ubiquity of probability distribution functions (PDF) with power law tails is still a matter of fascination and investigation in many scientific fields from linguistic, social, economic, computer sciences to essentially all natural sciences. In parallel, self-excited dynamics is a prevalent characteristic of many systems, from the physics of shot noise and intermittent processes, to seismicity, financial and social systems. Motivated by activation processes of the Arrhenius form, we bring the two threads together by introducing a general class of nonlinear self-excited point processes with fast-accelerating intensities as a function of ``tension''.  Solving the corresponding master equations, we find that a wide class of such nonlinear Hawkes processes have the PDF of their intensities described by a power law on the condition that (i)~the intensity is a fast-accelerating function of tension, (ii) the distribution of marks is two-sided with non-positive mean, and (iii) it has fast-decaying tails. In particular, Zipf's scaling is obtained in the limit where the average mark is vanishing. This unearths a novel mechanism for power laws including Zipf's law, providing a new understanding of their ubiquity.
\end{abstract}

\maketitle
\paragraph*{Introduction.}
	Many different types of data in the natural and social sciences exhibit power law density distributions of the size or frequencies of their characteristic variables.
	Namely, the probability densify function (PDF) $P(S)$ of a variable $S$ is given by $P(S) \sim 1/S^{1+\alpha}$ for large $S$ values, with $\alpha >0$. 
	Many mechanisms have been proposed to rationalise it~\cite{Powers1998,Sorbookcrit04,Newman05,SaiMalSor09}, such as proportional growth with additional conditions~\cite{MalSaiSor13}, family transformation of the Bose-Einstein distribution~\cite{HillWoodroofe75}, least-effort principles~\cite{Ferrer03}, optimisation between efficiency and faithfulness of self-reproduction~\cite{FuruKane03} and so on.

	Self-excited point processes assume that past events strongly influence the occurrence of future events. The Hawkes process~\cite{Hawkes1} is the simplest such process, where the intensity (probability per unit time that a new event occurs) is linear in the sum of the triggering influence of all past events. In the last decade, the Hawkes process and generalisations have enjoyed an explosive growth in the investigation of their properties and in a large set of applications in all fields of knowledge \cite{SorOsorio10,Tutorial17,Hawkes18,Reinhart18}. 
	
	Theoretically challenging, nonlinear self-excited processes have been scarcely investigated~\cite{BremaudMss96,GaoZhu18} except for a few special cases~\cite{QHawkesBouchaud}, even if they are a priori more suited to represent the interplay between stochasticity and nonlinear dynamics in many complex systems. Here, we study a class of nonlinear Hawkes processes characterised by fast-accelerating intensities as a function of an auxiliary field called the ``tension'', and report the first explicit solution that is applicable to a wide class of nonlinear Hawkes processes. We find that this class of nonlinear Hawkes family universally exhibits intensity distributions with power law tails. In particular, Zipf's scaling naturally appears when the distribution of marks is symmetric. A weaker condition is that the average mark is vanishing. These models are motivated by activation processes of the Arrhenius form, which are relevant in many applications in physics and also in seismicity and finance modelling as explained below.

\paragraph*{Model.}
	\begin{figure*}
		\centering
		\includegraphics[width=180mm]{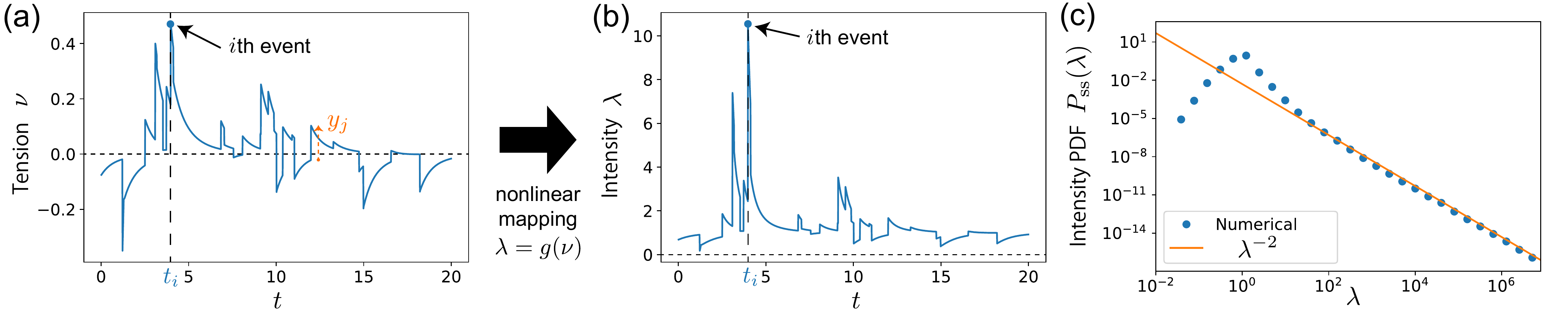}
		\caption{
			(a)~Sample tension trajectory $\{\nu(t)\}_t$ and (b)~the corresponding intensity trajectory $\{\lambda(t)\}_t$ generated by the nonlinear Hawkes process~\eqref{eq:tension-intensity-map} with (\ref{jh2bqbv}). The tension trajectory $\{\nu(t)\}_t$ exhibits random jumps with size $y_i$ distributed according to $\rho(y)$ and the corresponding intensity $\lambda$ is given by $\lambda(t)=g(\nu(t))$.
			(c)~Steady intensity distribution $P_{\mrss}(\lambda)$ for the exponential fast-accelerating intensity $g(\nu)\propto e^{\beta \nu}$ in the case where the mean mark is zero: $m=0$, exhibiting Zipf's law $\propto \lambda^{-2}$. These figures are based on  Monte Carlo simulations of the nonlinear Hawkes process with $\lambda=g(\nu)=\lambda_0 e^{\beta \nu}$, $h(t)=\sum_{k=1}^K \tilh_k e^{-t/\tau_k}$, and $\rho(y)=(1/\sqrt{2\pi \sigma^2})e^{-y^2/(2\sigma^2)}$ with $K=3$, $(\tau_1,\tau_2,\tau_3)=(1,0.5,2)$, $(\tilh_1,\tilh_2,\tilh_3)=(0.5,0.6,0.1)$, $\lambda_0=1$, $\beta=5$, and $\sigma=0.1$ (see Appendix for the detailed numerical scheme).
		}
		\label{fig:trajectory}
	\end{figure*}	
	
	The key ingredients of the nonlinear self-exciting Hawkes process considered here are the {\it intensity} $\lambda(t)$ and {\it tension} $\nu(t)$. Let us introduce a time series $\{t_i\}_{i}$, representing the timestamps of events, such as earthquakes, retweets on Twitter, or neural discharges in a brain. The intensity $\lambda (t)$ fully characterises the statistics of occurrence of events, such that an event occurs with probability $\lambda(t)dt$ during the interval $[t,t+dt)$. We assume that the intensity is a nonlinear positive and monotonically-increasing function of the system tension $\nu(t)$,
	\begin{equation}
		\lambda(t) = g(\nu(t))~,
		\label{eq:tension-intensity-map}
	\end{equation}
	where $g(\nu)$ is called the tension-intensity map. 
	The tension $\nu(t)$ quantifies the total stress due to historical events, such as resulting from elastic deformations of the crust
	induced by earthquakes. In finance, $\lambda(t)$ can represent the rate of volatility jumps and $\nu(t)$ is the rate of 
	financial returns whose amplitude exceeds some threshold.
	The tension at a given time is obtained as the sum of perturbations over all past events (see Fig.~\ref{fig:trajectory}(a) for a realisation), 
	such that 
	\begin{equation}
	\nu(t) = \sum_{i=1}^{N(t)}y_i h(t-t_i)~,
	\label{jh2bqbv}
	\end{equation}
	where each event $i$ has a mark $y_i$ distributed according to the PDF $\rho(y)$ and $N(t)$ is the number of events during $[0,t)$. Combining the relation~\eqref{eq:tension-intensity-map} between tension and intensity and (\ref{jh2bqbv}), we obtain a nonlinear version of the Hawkes process: $\lambda(t) = g\left(\sum_{i=1}^{N(t)}y_i h(t-t_i)\right)$. To represent that high tension promotes future events, we assume that the tension-intensity map is a non-decreasing function. For an affine function $g(\nu) = \nu_0 + \nu$, the model~(\ref{eq:tension-intensity-map}) with (\ref{jh2bqbv}) reduces to the original linear Hawkes process and $y_i$ can be interpreted as the average number of events of first generation triggered by event $i$ and is thus called the fertility of event $i$, by imposing $\int_0^\infty h(t)dt=1$. The memory function $h(t)\geq 0$ controls the distribution in time of the triggered events and decays to zero for large $t$.

\paragraph*{Conditions.}
	There is a large variety of nonlinear Hawkes processes defined via the pair of functions $g(\nu)$ and $\rho(y)$. Here we focus on the wide class of nonlinear Hawkes process that obey the three following conditions:
	\begin{enumerate}
		\item[(i)] the tension-intensity map $g(\nu)$ is a fast-accelerating intensity, defined to diverge faster than any second-order polynomial: $g(\nu) > O(\nu^2)$ for large $\nu$;
		\item[(ii)] the mark distribution is two-sided with nonpositive mean, such that $\int_{-\infty}^0 \rho(y)dy\neq 0$, $\int_0^{\infty} \rho(y)dy\neq 0$, and $m:=\int_{-\infty}^\infty y\rho(y)dy\leq 0$;
		\item[(iii)] the mark distribution has fast-decaying tails, such that $\Phi(x):=\int_{-\infty}^\infty dy\rho(y)(e^{xy}-1)$ exists. $\Phi(x)=0$ has only two roots, where one is zero and the other is $c^*\geq 0$.
	\end{enumerate}
	Here, the Bachmann-Landau-like inequality notation $a(x)>O(b(x))$ means $\lim_{x\to \infty}a(x)/b(x)=\infty$. Also, the condition~{(iii)} essentially means that fat-tail mark distributions, such as power law distributions, are out of scope in this Letter. Remarkably, all nonlinear Hawkes processes satisfying these three conditions have their steady-state intensity PDFs obeying the universal power law scaling, as we show below. Typical analytical forms satisfying condition (i) include $g(\nu)\propto \nu^{n}$ with $n>2$ and 
	\begin{equation}
		\label{eq:ExpIntensity}
		g(\nu) = \lambda_0e^{\beta \nu}~,
	\end{equation}
	which is motivated by the physics of rupture~\cite{SaiSor-And05} and earthquakes~\cite{OuilSor05_1,OuilSor05_2}, modelled as activated processes following an Arrhenius law. Indeed, we assume that the tension $\nu$ is proportional to the seismic energy $E$ (itself proportional to the total mechanical stress in the Earth crust), and let us assume that an earthquake happens if the system's state jumps over an energy barrier $E_0$ from a metastable state to another. According to the Arrhenius law, the escape rate is proportional to $e^{-\beta(E_0-E)}\propto e^{\beta \nu}$ with a disordered-enhanced effective inverse temperature $\sim \beta$ \cite{SaiSor-And05}, consistently with Eq.~\eqref{eq:ExpIntensity}. This exponential dependence~(\ref{eq:ExpIntensity}) also encompasses the class of multifractal processes emerging from 
	the interplay between exponential activation and long memory \cite{FiliSorMulti11}, which have been shown to be relevant to model financial volatility \cite{JiangRevMult19}. 
	By construction, the tension is dependent on all the marks of previous events, while future marks are drawn independently of the past history. This is consistent with the empirical unpredictability of earthquake magnitudes.
	
	Condition (ii) guarantees the stationarity of the model as a result of the cumulative contribution of the negative marks $y_i<0$, which prevent
	$\nu$ from diverging. An event with a negative (positive) mark $y_i$ is likely to inhibit (induce) future events. The coexistence of events that inhibit and of events that promote future activity in our nonlinear Hawkes model is a fundamental extension to the general class of Hawkes processes. This allows us to realistically account for ubiquitous inhibitory effect in real complex systems, such as the random mechanical stress-relaxation after earthquake in seismology, or inhibitory synaptic potentials in neural networks. Note that, in contrast, the standard Hawkes process and many other versions only have positive marks, corresponding to taking into account excitations exclusively.
	
\paragraph*{Power law intensity PDF.}
	When conditions (i)-(iii) are satisfied, the steady-state PDF of the intensity $\lambda$ is analytically given by
	\begin{equation}
		\label{eq:gen_solution}
		P_{\mrss}(\lambda) \propto \lambda^{-1}\left[e^{-a\nu}\left\{\frac{dg}{d\nu}\right\}^{-1}\right]_{\nu=g^{-1}(\lambda)}, \>\>\> a:= \frac{c^*}{h(0)}
	\end{equation}
	with $c^*$ being the nonnegative root of $\Phi(c^*)=0$ (see SM). This formula readily reduces to various power law asymptotic forms, such as 
	\begin{equation}
		\label{eq:powerlaw_intensity}
		P_{\mrss}(\lambda) \propto 
		\begin{cases}
			\lambda^{-2-\beta^{-1}a} & (\mbox{for $g(\nu) \simeq \lambda_0 e^{\beta \nu}$, $\beta>0$}) \\
			\lambda^{-2+\frac{1}{n}}e^{-a\left(\frac{\lambda}{\lambda_0}\right)^{\frac{1}{n}}} & 	(\mbox{for } g(\nu)\simeq \lambda_0\nu^n, \>\> n>2).
		\end{cases}
	\end{equation}
	Beyond conditions (i)-(iii), no other properties or details, including the shape of the memory function, change the robust classes given by expressions~\eqref{eq:powerlaw_intensity}. 

\paragraph*{Zipf's law.}
	Result (\ref{eq:powerlaw_intensity}) implies that Zipf's scaling appears as an important subclass of the non-linear Hawkes processes as a special case $a=0$:   
	\begin{equation}
		\label{eq:Zipf_intensity}
		P_{\mrss}(\lambda) \propto 
		\begin{cases}
			\lambda^{-2} &  	(\mbox{for } g(\nu)>O(\nu^n) \mbox{ for any }n) \\
			\lambda^{-2+\frac{1}{n}} & 	(\mbox{for } g(\nu)\simeq \lambda_0\nu^n, \>\> n>2)~,
		\end{cases}
	\end{equation}
	except for minor logarithmic corrections. The condition $a=0$ is realised exactly when
	$c^*=0$, which holds for zero-mean mark $m=0$, implying $a=c^*/h(0)=0$. This is for instance realised for symmetric mark distribution $\rho(y)=\rho(-y)$. Approximate Zipf's distributions are obtained when $a=c^*/h(0)$ is small, which occurs for large $h(0)$.
	
	Symmetric mark distributions are realised in the physics of earthquakes as discussed in \cite{OuilSor05_2}. Indeed, the stress perturbations induced by a (small) earthquake correspond to the stress field of a double-couple, which can be simply represented by a concentrated set of four forces of the same norm, summing to zero (zero total force) and with total torque also equal to zero. A large earthquake is just a set of double-couple sources places along its fault surface. The stress induced by a double-couple has a nice butterfly symmetry with four lobes, two positive and two negative ones, and is perfectly symmetric. With the correspondence that the tension $\nu$ is proportional to stress, and is given by (\ref{jh2bqbv}), and that the exponential intensity function (\ref{eq:ExpIntensity}) derives from the physics of earthquake nucleation with Arrhenius law with an effective temperature \cite{OuilSor05_2}, this justifies the symmetric property of the distribution of marks for earthquakes. 
 
	The results of numerical simulations for $m=0$ (zero mean marks) are presented in Fig.~\ref{fig:trajectory}. Panel a) shows a typical realisation of $\nu$ for case~(\ref{eq:ExpIntensity}), while panel b) shows the derived temporal evolution of $\lambda$. Panel c) shows the corresponding steady-state PDF of $\lambda$ obeying Zipf's law (see also SM for numerical simulations for the negative-mean cases $m<0$).

\paragraph*{Field-master equation.}
	\begin{figure*}
		\centering
		\includegraphics[width=180mm]{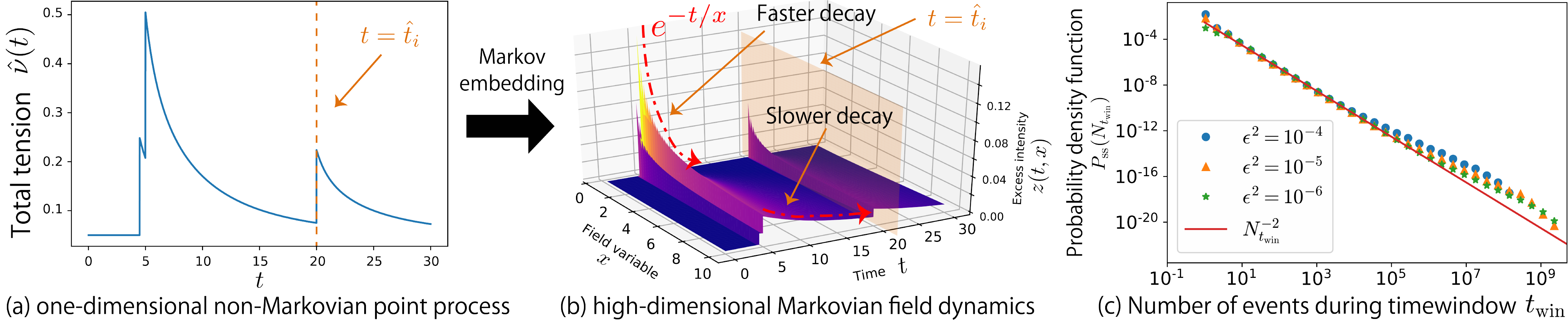}
		\caption	{
					(a-b)~Schematic of the Markov embedding from the one-dimensional non-Markovian process $\nu(t)$ to the infinite-dimensional Markovian field dynamics $\{z(t,x)\}_{x \in \bm{R}^+}$. The original process~{(a)} is non-Markovian because its time evolution (\ref{eq:tension-intensity-map}) with (\ref{jh2bqbv}) depends on all the history $\{\nu(s)\}_{s\leq t}$. On the other hand, the field dynamics~{(b)} is Markovian because its time evolution~\eqref{eq:SPDE} depends only on the current configuration of the field variable $\{z(t,x)\}_{x\in \bm{R}^+}$. Note that the auxiliary field variable $x\in \bm{R}^+$ is introduced according to Eqs.~\eqref{eq:decomposition_of_h} and~\eqref{eq:SPDE} and is interpreted as a ``position'' at which the field is evaluated. The decay speed is faster for smaller $x$, while it is slower for larger $x$ according to Eq.~\eqref{eq:SPDE}. (c)~Statistics for the number of events $N_{t_\mrwin}$ occurring in short time windows of size $t_\mrwin$ in the diffusive scaling limit~\eqref{eq:scaling_diffusive} for various $\eps$. We observe a Zipf law up to the upper cut-off $N_{\mathrm{cut}}=O(\eps^2)$. Beyond the cutoff, a fatter tail is numerically observed (see SM). 
				}
		\label{fig:MarkovEmbedding}
	\end{figure*}
	Our general result for a large class of memory functions can be derived using our recently introduced field-master-equation framework~\cite{KS_PRL2020,KS_PRR2020} (see also Appendix for the technical detail). The main idea is to convert the original low-dimensional non-Markovian stochastic process onto a high-dimensional Markovian field dynamics. This technique is called Markov embedding and has been applied for some special cases, such as memory functions composed of discrete sums of exponentials (see Refs.~\cite{Zwanzig,Kupferman} for the generalised Langevin equation and Refs.~\cite{ME_Hawkes1,ME_Hawkes2} for Hawkes processes).

	The Markov embedding scheme can be formulated for the nonlinear Hawkes process (\ref{eq:tension-intensity-map}) with (\ref{jh2bqbv}) as follows.
	Let us decompose the memory kernel $h(t)$ as a continuous sum of exponentials. This amounts to representing $h(t)$ 
	as a Laplace-like transform of another function ${\tilde h}(x)$
	of the auxiliary variable $x \in (0,\infty)$:
	\begin{equation}
		\label{eq:decomposition_of_h}
		h(t) = \int_0^\infty dx ~\tilh(x)  e^{-t/x}.
	\end{equation}
	Based on this decomposition, the original process (\ref{eq:tension-intensity-map}) with (\ref{jh2bqbv}) is equivalent to a Markovian stochastic partial differential equation (SPDE) for the excess tension $\{z(t,x)\}_{x\in \bm{R}^+}$
	\begin{equation}
		{\partial z(t,x) \over \partial t} = - {z(t,x) \over x} + \tilh(x) \xi^{\mathrm{P}}_{\rho(y);\lambda(t)}
		\label{eq:SPDE}
	\end{equation}
	with the total tension $\nu(t)= \int_0^\infty dxz(t,x)$ (see Figs.~\ref{fig:MarkovEmbedding}a and \ref{fig:MarkovEmbedding}b for schematics of the Markov embedding scheme) and the compound Poisson noise $\xi^{\mathrm{P}}_{\rho(y);\lambda(t)} = \sum_{i=1}^{N(t)}y_i\delta(t-t_i)$. Remarkably, while the original process is non-Markovian in a one-dimensional space $\nu(t)$, the field dynamics is Markovian in the infinite-dimensional space $\{z(t,x)\}_{x\in \bm{R}^+}$. 
	
	The equivalence between the original nonlinear Hawkes process (\ref{eq:tension-intensity-map}) with (\ref{jh2bqbv}) and the SPDE~\eqref{eq:SPDE} can be shown as follows: the formal solution of the SPDE~\eqref{eq:SPDE} is given by $z(t,x)=\int_{-\infty}^t\tilh(x)e^{-(t-s)/x}\xi^{\mathrm{P}}_{\rho(y);\lambda}(s) ds = \sum_{i=1}^{N(t)}\tilh(x) y_ie^{-(t-t_i)/x}$. The total tension is then given by $\nu(t) := \int_0^\infty dxz(t,x) = \sum_{i=1}^{N(t)}y_i \int_0^\infty dx \tilh(x)e^{-(t-t_i)/x} = \sum_{i=1}^{N(t)}y_i h(t-t_i)$. This is equivalent to (\ref{jh2bqbv}). We thus find that the Markovian SDE~\eqref{eq:SPDE} is a correct representation after Markov embedding.

	The SPDE~\eqref{eq:SPDE} can be regarded as the ``physical dynamics'' of the field variable $\{z(t,x)\}_{x\in \bm{R}^+}$, since $x$ can be considered as the ``physical position'' in  $\bm{R}^+:=(0,\infty)$ on which the field is evaluated. This interpretation has the advantage that the functional methods for various SPDEs of stochastic field dynamics are available for advanced analytics (e.g., the functional Fokker-Planck equations for the reaction-diffusion equations~\cite{GardinerB}). 
	
	Since the SPDE~\eqref{eq:SPDE} is Markovian, we can obtain the corresponding master equation. By introducing the probability density functional (PDF) $P_t[z]:=P_t[\{z(t,x)\}_{\bm{R}^+}]$,
	the field master equation is given by
	\begin{align}
		{\partial P_t[z] \over \partial t} &= \left(\mathcal{L}_{\mathrm{A}} + \mathcal{L}_{\mathrm{J}} \right)P_t[z]
		\label{eq:field_master_equation}
	\end{align}
	with the advective and jump Liouville operators $\mathcal{L}_{\mathrm{A}}$ and $\mathcal{L}_{\mathrm{J}}$, respectively, defined by
	\begin{subequations}
		\begin{align}
			\mathcal{L}_{\mathrm{A}}P_t &:= \!\! \int_0^\infty dx {\delta \over \delta z(x)} {z(x) \over x}P_t[z] \label{eq:field_master_adj} \\
			\mathcal{L}_{\mathrm{J}}P_t &:= \!\! \int_{-\infty}^\infty \!\!\!\!\! dy \rho(y)G[z-y \tilde{h}]P_t[z-y \tilde{h}]-G[z]P_t[z]
		\end{align}
	\end{subequations}
	with $G[z]:=g(\int_0^\infty z(t,x)dx)$ and $\rho(y)$ is the mark distribution. 
	
	Note that $P_t[z]$ is a path probability measure: the probability is given by $P_t[z]\mathcal{D}z$ that the configuration of the field variable is nearly-equal to $\{z(t,x)\}_{\bm{R}^+}$, 
	where $\mathcal{D}z := \prod_{x\in \bm{R}^+}dz(x)$ is the path-integral volume element. In addition, the ensemble average $\la A \ra$ is given by the path integral $\la A\ra := \int A P_t[z]\mathcal{D}z$. Technically, the field master equation~\eqref{eq:field_master_equation} should be interpreted as a formal limit from discrete underlying descriptions according to the standard convention (see Ref.~\cite{GardinerB} and Appendix). The steady-state solution $P_{\mrss}[z]$ is related to the steady-state intensity PDF $P_{\mrss}(\lambda)$ as $P_{\mrss}(\lambda) = \int \mathcal{D}z \delta\left(\lambda-\int_0^\infty dxz(t,x)\right)P_{\mrss}[z]$.
	
\paragraph*{Derivation.}
	We now provide an outline of the theoretical derivation of the solution of the field-master equation (see Appendix for more detailed calculations and an illustrative-case study with the exponential memory). Let us introduce $\phi[z]:=G[z]P_{\mrss}[z]$ to rewrite Eq.~\eqref{eq:field_master_equation} in the steady state as
	\begin{equation}
		0 = \int_0^\infty dx\frac{\delta}{\delta z} \left(\frac{z\phi[z]}{x G[z]}\right) + \int_{-\infty}^\infty dy\rho(y)\phi[z-y\tilh] -\phi[z].
	\end{equation}
	Since the first term is negligible for large $z$ assuming the condition~{(i)}, the asymptotic solution satisfies 
	\begin{equation}
		\int_{-\infty}^\infty dy\rho(y)\phi[z-y\tilh] -\phi[z]\approx 0 \>\>\> \mbox{for large }z.
		\label{eq:outlineDer_trans1}
	\end{equation}
	Under conditions~{(ii)} and (iii), its solution is given by $\phi[z]\approx C_0[Z']e^{-c^*W}$ with $W:=z(x^*)/h(x^*)$ and $Z'(x):=z(x)-h(x)z(x^*)/h(x^*)$ for $\bm{R}'^{+}:=\bm{R}^+\setminus \{x^*\}$ by selecting an appropriate number $x^*\in \bm{R}$. Here $C_0$ is an arbitrary functional without $W$ as an argument. After a path integral for marginalisation, we obtain 
	\begin{equation}
		P_{\mrss}(\nu):= \int \mathcal{D}z P_{\mrss}[z]\delta\left(\nu-\int_0^\infty dxz(x)\right) \approx \frac{e^{-c^*\nu/h(0)}}{g(\nu)}.
	\end{equation}
	Equation~\eqref{eq:gen_solution} then follows by the change of variable $\nu\to\lambda$.

\paragraph*{Intuition.}
	Let us consider the case of an exponential growing intensity (\ref{eq:ExpIntensity}) and a simple exponential memory kernel $h(t)=(n/\tau)e^{-t/\tau}$, 
	where the integral of the memory, $n=\int_0^\infty h(t)dt>0$, would be interpreted as the branching ratio in the linear case. 
	Suppose that the initial tension is zero and thus the initial intensity is $\lambda_0$. Naively, one could infer that the typical 
	waiting time till the next event, the expected event interval (EEI), is given by ${1 \over \lambda_0}$. Choosing the parameters such that $\tau \ll {1 \over \lambda_0}$
	 would imply that the influence of an event in triggering future events is extremely localised temporally and one should expect no intermittency, no power laws and a rather trivial behaviour. This reasoning is wrong as it neglects the nonlinear nature of the model with strong feedback loops. Indeed, defining the small parameter $\eta := \lambda_0 \tau \ll 1$, after one event occurs with positive mark $y>0$, the intensity is given instantaneously by $\lambda(t)=\lambda_0 e^{\beta \lambda_0 y n/\eta}$ and the corresponding EEI is of the order of $\frac{1}{\lambda(t)}={1 \over \lambda_0} e^{-\beta \lambda_0 y n/\eta}$. Paradoxically, as the memory $\tau$ of the event is vanishingly smaller than the naive characteristic time scale ${1 \over \lambda_0}$, the time needed for the next event to be triggered becomes exceedingly smaller, since ${1 \over \lambda_0} e^{-\beta \lambda_0 y n/\eta}  \ll \tau \ll {1 \over \lambda_0}$ for sufficiently small $\eta$ such that $\eta  \ln {1 \over \eta}  \ll \beta \lambda_0 y n$. Hence, in contradiction with the naive view, a very short memory enhances triggering and creates a very rich bursty dynamics of events. Readily generalised to multiple events, this reasoning gives an intuition on the basic source of the scale-free nature of the power-law intensity PDF~\eqref{eq:powerlaw_intensity}, suggesting the absence of both characteristic intensity and EEI. 
	
\paragraph*{Number-of-events statistics.}
	The intensity PDF is a fundamental quantity to characterise temporal properties of point processes and 
	allows one to derive various other quantities. One such variable that is directly observable is the total number of events $N_{t_\mrwin}$ 
	occurring in a finite time window $t_\mrwin$. Assuming symmetric mark distributions $\rho(y)=\rho(-y)$, we show that Zipf's law also holds for the distribution of $N_{t_\mrwin}$.
	For simplicity, we consider the diffusive scaling limit (i.e., essentially equivalent to the system-size expansion~\cite{GardinerB}, an established perturbative method invented by van Kampen~\cite{VanKampen} based on realistic scaling assumptions; see SM for a brief review) by introducing a small parameter $\eps$: 
	\begin{equation}
		g(\nu) = \frac{1}{\eps^2}\bar{g}(\nu), \>\>\> 
		\rho(y) = \frac{1}{\eps}\bar{\rho}\left(\frac{y}{\eps}\right)
		\label{eq:scaling_diffusive}
	\end{equation}
	with $\eps$-independent functions $\bar{g}$ and $\bar{\rho}$. We focus on the case with $\bar{g}(\nu)=\lambda_0e^{\beta \nu}$. In this diffusive limit, corresponding to the mark size being typically much smaller in absolute value than the tension at any given time, the statistics of $N_{t_\mrwin}$ obeys Zipf's law for a sufficiently short time window $t_\mrwin$ (see Fig.~\ref{fig:MarkovEmbedding}c):
	\begin{equation}
		P_{\mrss}(N_{t_\mrwin})\propto N_{t_\mrwin}^{-2} \>\>\> \mbox{for } N_{t_\mrwin} < N_{\mathrm{cut}}
		\label{eq:Zipf_eventNumbers}
	\end{equation}
	as an intermediate asymptotics~\cite{BarenblattB} with upper cutoff $N_{\mathrm{cut}}=O(\eps^{-2})$.
	
	This relation can be derived from a superposition of Poisson statistics. Let us consider a long time series in $[0,T)$ and then randomly select a timepoint $\tau \in [0,T)$. For a sufficiently-short time window $[\tau,\tau+t_{\mrwin})$, we can assume that $\lambda(t)$ is constant and the number of events obeys the Poisson statistics $P(N_{t_\mrwin} | \lambda)=(\lambda t_{\mrwin})^{N_{t_\mrwin}}e^{-\lambda t_{\mrwin}}/N_{t_\mrwin}!$. Choosing $\tau$ randomly and neglecting dependences between count numbers across different windows, the unconditional distribution is given by the superposition of the Poisson distribution as
	\begin{equation}
		P_{\mrss}(N_{t_\mrwin}) \simeq \int P(N_{t_\mrwin} | \lambda)P_{\mrss}(\lambda)d\lambda \propto N_{t_\mrwin}^{-2}.
	\end{equation}
	This examples shows that Zipf's law~\eqref{eq:Zipf_intensity} for the intensity PDF $P_{\mrss}(\lambda)$ is directly relevant to Zipf's laws for other observable quantities.
	
	Theoretically, relation~\eqref{eq:Zipf_eventNumbers} is expected to hold only up to the cutoff $N_{\mathrm{cut}}$ (see Appendix), which diverges as $\eps \to 0$, guaranteeing the robust universality of Zipf's law for $P_{\mrss}(N_{t_\mrwin})$ in the diffusive limit. Beyond the cutoff, we numerically observe a fatter tail stemming from dependences between count numbers in adjacent time windows, which becomes dominant at very high count numbers, as can be seen from its impact on $\nu$ given by \eqref{jh2bqbv}.
	
\paragraph*{Conclusion.}	
	As power laws are widely observed in many complex systems, our theoretical finding suggests the nonlinear self-excited mechanism as an explanation for the universality of power laws. Intuitively, these properties emerge from the intricate interplay between a kind of multiplicative process, memory and endogeneity / reflexity. Our new tools and results will be useful for data analysis of real complex systems. Interested readers are referred to Ref.~\cite{KanazawaPRR2021} for more mathematical details.

\begin{acknowledgments}
	This work was supported by (i) JST, PRESTO Grant Number JPMJPR20M2, Japan, (ii) the Japan Society for the Promotion of Science KAKENHI (Grant No.~20H05526), (iii) Intramural Research Promotion Program in the University of Tsukuba and (iv) partially by the National Natural Science Foundation of China under grant No. U2039202. The numerical computation for Fig.~\ref{fig:MarkovEmbedding}c was carried out at the Yukawa Institute Computer Facility.
\end{acknowledgments}

\begin{widetext}
\appendix

\section{Methods}
\subsection{Markov embedding (discrete sum of exponentials)}
	Let us first focus on the case of a superposition of exponentials:
	\begin{equation}
		h(t) = \sum_{k=1}^K \tilh_k e^{-t/\tau_k}.
		\label{SIeq:superpositionOfExpon}
	\end{equation}
	For this case, Eq.~\eqref{jh2bqbv} can be converted into the following Markovian dynamics, 
	\begin{equation}
		\nu(t) = \sum_{k=1}^K z_k(t), \>\>\> {dz_k \over dt} = -{z_k \over \tau_k} + \tilh_k\xi^{\rm{P}}_{\rho(y); \lambda(t)}
		\label{SIeq:SDE_Markov_discrete}
	\end{equation}
	with the state-dependent Poisson noise $\xi_{\rho(y);\lambda(t)}^{\rm{P}}$, defined by
	\begin{equation}
		\xi_{\rho(y);\lambda(t)}^{\rm P}=\sum_{i=1}^{N(t)}{y_i\delta(t-t_i)},
	\end{equation}
	where $t_i$ is the $i$th event time and $y_i$ is a random number obeying a given distribution $\rho\left(y\right)$. 
	Note that the probability that an event occurs within interval $[t,t+dt)$ is given by
	\begin{equation}
		\lambda (t)dt=g(\nu(t))dt.
	\end{equation}
	This technique~\cite{ME_Hawkes1,ME_Hawkes2} is called Markov embedding, where low-dimensional non-Markovian dynamics is converted onto higher-dimensional Markovian dynamics. 
	 
	We note that this Markov embedding framework is sufficiently general since any memory kernel $h(t)$ can be written 
	as a continuous sum of exponentials (via the Laplace transformation), which can be approximated by the discrete-sum formula~\eqref{SIeq:superpositionOfExpon}, such that 
	\begin{equation}
		h(t)=\int_{0}^{\infty}dx\tilh (x)e^{-t/x}\approx \sum_{k=1}^{K}\tilh_k e^{-t/\tau_k}.
	\end{equation}
	In this sense, the discrete representation $\tilh_k$ corresponds to the continuous function representation $\tilh(x)$ via this relationship. This method can be formally generalised for the general continuous sum of exponentials as shown in Appendix~\ref{sec:SI_continuousSum}. 

\subsection{Numerical scheme}
	We have numerically studied Eq.~\eqref{jh2bqbv} based on the Monte Carlo simulations of Eq.~\eqref{SIeq:SDE_Markov_discrete} for Fig.~\ref{fig:trajectory}.  
	Let us introduce a discretised time series,
	\begin{equation}
		0=s_0<s_1<\dots<s_N=T,\ \ \Delta s_i=s_{i+1}-s_i. 
	\end{equation}
	Equation~\eqref{SIeq:SDE_Markov_discrete} reads 
	\begin{equation}
		z_k(s_{i+1}) - z_k(s_i) = - {z_k(s_i) \over \tau_k} \Delta s_i + 
		\begin{cases}
			0 & (\mbox{Probability}=1-\lambda(s_i)\Delta s_i) \\
			\tilh_k y_k & (\mbox{Probability}=\lambda(s_i)\Delta s_i)
		\end{cases}
		\label{SIeq:NumericalScheme}
	\end{equation}
	for $i=0,\dots,N-1$. The mark sequence $\left\{y_k\right\}_k$ obeys the normal distribution
	\begin{equation}
		P(y_k)={1 \over \sqrt{2\pi\sigma^2}}e^{-(y_k-m)^2/(2\sigma^2)}
	\end{equation}
	with mean $m$ and variance $\sigma^2$. 
	The time step $\Delta s_k$ in Eq.~\eqref{SIeq:NumericalScheme} must be sufficiently small, such that $\lambda(s_i) \Delta s_i\ll1$. 
	We therefore employ an adaptive scheme
	\begin{equation}
		\Delta s_i=\min{\left\{\Delta t_{\max}^{(1)},\frac{\Delta t_{\max}^{(2)}}{\lambda(s_i)}\right\}}
	\end{equation}
	with $\Delta t_{\max}^{(1)}$ and $\Delta t_{\max}^{(2)}$. In addition, we introduce a finite cutoff for the tension-intensity map,
	\begin{equation}
		\lambda(t)=g(\nu(t))=\min{\left\{\lambda_0e^{\beta\nu(t)},\lambda_{\max}\right\}}
	\end{equation}
	with $\lambda_{\max}={10}^7$, to control rounding error.

	For the numerical trajectory generated by Eq.~\eqref{SIeq:NumericalScheme}, we obtain the empirical steady intensity distribution as 
	\begin{equation}
		P_{\mathrm{ss}}(\lambda)=\la \delta(\lambda - \lambda(t)) \ra = \lim_{T\to \infty} {1\over T} \int_0^T\delta(\lambda - g(\nu(s)))ds \approx {1\over T} \sum_{i=0}^{N-1}\delta(\lambda - g(\nu(s_i)))\Delta s_i
	\end{equation}
	under the assumption of ergodicity. In addition, we have applied a parallel computing method,
	\begin{equation}
		P_{\mathrm{ss}}(\lambda) = \left<\lim_{T\to \infty} {1\over T} \int_0^T\delta(\lambda - g(\nu(s)))ds\right> 
		\approx {1\over N_{\mathrm{PC}}}\sum_{j=1}^{N_{\mathrm{PC}}}{1\over T} \sum_{i=0}^{N-1}\delta(\lambda - g(\nu(s_i)))\Delta s_i
	\end{equation}
	with a total number $N_{\mathrm{PC}}$ of parallel threads.

	\subsubsection{Numerical simulation for the negative mean mark case $m<0$}
		\begin{figure}
			\centering
			\includegraphics[width=180mm]{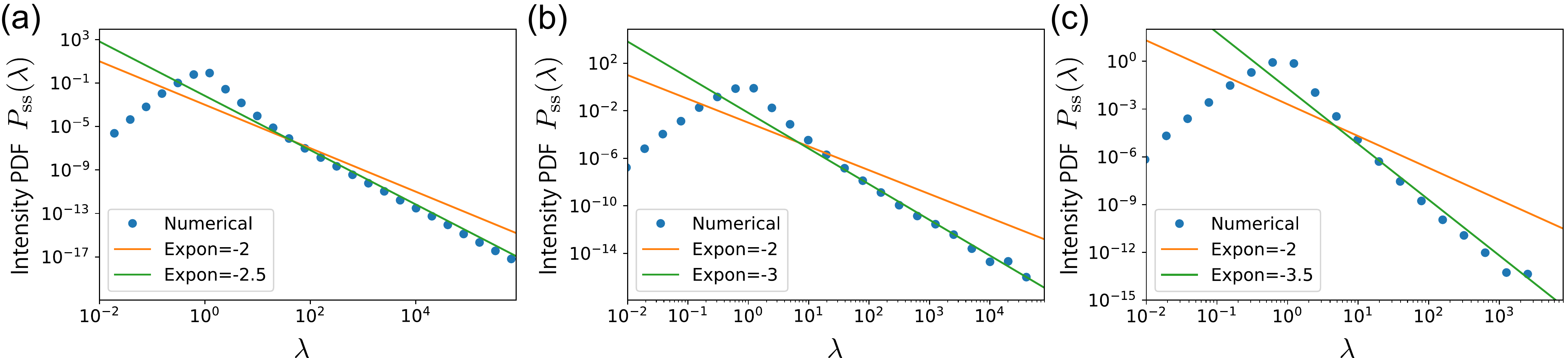}
			\caption{
					Numerical simulations for the negative mean mark cases, where the deviations from Zipf's scaling are observed, such that $P_{\mrss}(\lambda)\propto \lambda^{-2-a}$ with (a)~$a=0.5$, (b)~$a=1$, and (c)~$a=1.5$. The parameters are summarised in Table~\ref{SItable_param}. 
			}
			\label{SIfig:AsymmetricMarks}
		\end{figure}
		Since the PDF obeying Zipf's law is shown for the zero mean mark case in the main text as Fig.~\ref{fig:trajectory}c, here we provide numerical simulations for the cases with negative mean marks $m\leq 0$, where the exponent deviates from Zipf's scalings. Our theory predicts the relation 
		\begin{equation}
			P_{\mrss}(\lambda) \propto \lambda^{-2-\beta^{-1}a}, \>\>\> a = -\frac{2m}{\sigma^2 h(0)}
		\end{equation}
		for the Gaussian mark distribution $\rho(y)=e^{-(y-m)^2/(2\sigma^2)}/\sqrt{2\pi\sigma^2}$ (see Appendix~\ref{app:sec:Gaussian_mark} for the derivation), which agrees with the numerical intensity PDFs as shown in Fig.~\ref{SIfig:AsymmetricMarks}. 

	\begin{table}
		\begin{tabular}{|c||c|c|c|c|}
		\hline
		parameter                     & Figure 1c      & Figure~\ref{SIfig:AsymmetricMarks}a      & Figure~\ref{SIfig:AsymmetricMarks}b      & Figure~\ref{SIfig:AsymmetricMarks}c    \\ \hline
		$\Delta t_{\max}^{(1)}$       & \multicolumn{4}{c|}{$0.1$}                              \\ \hline
		$\Delta t_{\max}^{(2)}$                       &  \multicolumn{4}{c|}{$0.01$}                              \\ \hline
		$K$                           & \multicolumn{4}{c|}{$3$}                                               \\ \hline
		$\{\tau_i\}_{i=1,\dots,K}$    & \multicolumn{4}{c|}{$(1.0,0.5,2.0)$}                                   \\ \hline
		$\{\tilde{h}_i\}_{i=1,\dots,K}$ & \multicolumn{4}{c|}{$(0.5,0.6,0.1)$}                                   \\ \hline
		$\lambda_0$                   & \multicolumn{4}{c|}{$1$}                                               \\ \hline
		$a$                       		& $0$ (Zipf)     & $0.5$ & $1$ & $1.5$\\ \hline
		$\beta$                       & $5$            & \multicolumn{3}{c|}{$1$}                              \\ \hline
		$m$                           & $0$            & \multicolumn{3}{c|}{$\displaystyle -\frac{a\sigma^2 h(0)}{2}$} \\ \hline
		$\sigma$                      & $0.1$          & \multicolumn{3}{c|}{$0.5$}                            \\ \hline
		$T$                           & $5\times 10^4$ & \multicolumn{3}{c|}{$5\times 10^6$}                   \\ \hline
		$N_{\rm PC}$                  & \multicolumn{4}{c|}{$8$}                                               \\ \hline
		\end{tabular}
		\caption{
			Summary table of the parameters used in the numerical simulations. 
		}
		\label{SItable_param}
	\end{table}

	\subsubsection{Parameters}
		In our work, we set the parameters summarised in Table~\ref{SItable_param} for numerical simulations. 

\section{Master equation}
	Since Eq.~\eqref{SIeq:SDE_Markov_discrete} is Markovian, we can derive the corresponding master equation (i.e., the time-evolution equation for the probability density function (PDF)). 
	Let us consider the phase point $\bm{z}:=(z_1,\dots ,z_K)$ and its PDF $P_t(\bm{z})$. Indeed, the PDF satisfies the following master equation 
	\begin{equation}
		\frac{\partial P_t(\bm{z})}{\partial t}=\sum_{k=1}^{K} \frac{\partial}{\partial z_k}\frac{z_k}{\tau_k}P_t(\bm{z})
		+\int_{-\infty}^{\infty}dy\rho(y)\left\{G(\bm{z}-y\bm{\tilh})P_t(\bm{z}-y\bm{\tilh})-G(\bm{z})P_t(\bm{z})\right\}
		\label{SIeq:master_discrete}
	\end{equation}
	where we introduce $G(\bm{z}):= g\left(\sum_{k=1}^Kz_k\right)$ and $\bm{\tilh}:=\left(\tilh_1, \dots, \tilh_K \right)$. 
	In the following, we focus on the case with non-positive mean mark:
	\begin{equation}
		m:= \int_{-\infty}^\infty y\rho(y)dy \leq 0.
	\end{equation}
	Remarkably, the solution explodes for $m>0$ (see Appendix~\ref{sec:SM:existenceSol} for an intuitve discussion on the condition of the explosive solutions). 

	\subsection{Derivation}\label{sec:SM:der_masterEq.}
		Equation~\eqref{SIeq:master_discrete} can be derived as follows. Let us introduce an arbitrary function $f(\bm{z})$. The time-evolution of $f(\bm{z})$ is given by
		\begin{equation}
			df(\bm{z}) = 	\begin{cases}
								-\sum_{k=1}^K {z_k \over \tau_k} {\partial f(\bm{z}) \over \partial z_k} dt, & (\mbox{No jump during }[t,t+dt):\mbox{prob.}=1-\lambda(t)dt) \\
								f(\bm{z}+y\bm{\tilh}) - f(\bm{z}), & (\mbox{Jump in }[t,t+dt): \mbox{prob.}=\lambda(t)\rho(y)dtdy)
							\end{cases}
		\end{equation}
		with jump size $y$ obeying a given PDF $\rho(y)$. We take an ensemble average to obtain
		\begin{equation}
			\Bigg<{df \over dt}\Bigg> = \Bigg< -\sum_{k=1}^K {z_k \over \tau_k} {\partial f(\bm{z}) \over \partial z_k} +\int_{-\infty}^{\infty}dy\rho(y) G(\bm{z})\left[f(\bm{z}+y\bm{\tilh}) - f(\bm{z})\right] \Bigg>.
		\end{equation}
		Here we integrate by part to obtain 
		\begin{equation}
			-\int_{-\infty}^\infty d\bm{z} P_t(\bm{z}){z_k \over \tau_k} {\partial f(\bm{z}) \over \partial z_k}
			= \int_{-\infty}^\infty d\bm{z} f(\bm{z}){\partial \over \partial z_k} {z_k \over \tau_k}P_t(\bm{z}).
		\end{equation}
		We also apply a variable transformation $\bm{z}+y\bm{\tilh} \to \bm{z}$ to obtain
		\begin{equation}
			\int_{-\infty}^\infty d\bm{z} P_t(\bm{z}) G(\bm{z})f(\bm{z}+y\bm{\tilh})
			= \int_{-\infty}^\infty d\bm{z} f(\bm{z}) G(\bm{z}-y\bm{\tilh})P_t(\bm{z}-y\bm{\tilh}).
		\end{equation}
		This yields the identity
		\begin{equation}
			\int_{-\infty}^{\infty}d\bm{z} f(\bm{z})\left\{\frac{\partial P_t(\bm{z})}{\partial t}-\sum_{k=1}^{K} \frac{\partial}{\partial z_k}{\frac{z_k}{\tau_k} P_t(\bm{z})}-\int_{-\infty}^{\infty}dy\rho(y)\left[P_t(\bm{z}-y\bm{\tilh})G(\bm{z}-y\bm{\tilh})-P_t(\bm{z})G(\bm{z})\right]\right\}=0.
		\end{equation}
		Since this identity holds for an arbitrary function $f(\bm{z})$, we obtain Eq.~\eqref{SIeq:master_discrete}.

	\subsection{Solution for zero-mean mark distributions}\label{SIsec:sol_zero-mean}
		The asymptotic solution of the master equation~\eqref{SIeq:master_discrete} can be obtained as follows. 
		Let us define the steady PDF $P_{\mathrm{ss}}(\bm{z}):=\lim_{t\to \infty}P_t(\bm{z})$ and $\phi(\bm{z}):=G(\bm{z})P_{\mathrm{ss}}(\bm{z})$ to rewrite Eq.~\eqref{SIeq:master_discrete} as 
		\begin{equation}
			\sum_{k=1}^{K}{ {1 \over \tau_k} {\partial \over \partial z_k}\left(\frac{z_k}{G(\bm{z})}\phi(\bm{z})\right)}+\int_{-\infty}^{\infty}dy\rho(y)\phi\left(\bm{z}-y\bm{\tilh}\right)-\phi(\bm{z})=0
			\label{SIeq:master_zeroMean}
		\end{equation}	
		for $t\rightarrow\infty$. For large $\bm{z}$, the first term is negligibly small for the fast-accelerating intensity $G(\bm{z})\gg \left(\sum_{k=1}^{K}z_k\right)^2$. 
		We thus obtain
		\begin{equation}
			\int_{-\infty}^\infty dy\rho(y)\phi(\bm{z}-y\bm{\tilh}) - \phi(\bm{z}) \approx 0 \>\>\> \mbox{for large }\bm{z}.
			\label{SIeq:asymptoticEq_sfda}
		\end{equation}
		Let us apply a variable transformation from $\bm{z}=(z_1,\dots,\ z_K)$ to $\bm{Z}:=(W, Z_2, \dots Z_K)$ with 
		\begin{equation}
			z_1=\tilh_1 W,\>\>\> z_2=\tilh_2 W+Z_2,\>\>\> z_3=\tilh_3 W+Z_3,\>\>\> \dots \>\>\>,  z_K=\tilh_K W+Z_K,
			\label{SIeq:var_trans_intEq}
		\end{equation}
		which leads to
		\begin{equation}
			\phi(\bm{z}-y\bm{\tilh}) = \phi\left(\tilh_1(W-y), \tilh_2(W-y)+Z_2,\dots, \tilh_K(W-y)+Z_k\right).
		\end{equation}
		By defining 
		\begin{equation}
			\psi(W-y;Z_2,\dots, Z_K) = \psi(W-y;\bm{Z}'):= \phi\left(\tilh_1(W-y), \tilh_2(W-y)+Z_2,\dots, \tilh_K(W-y)+Z_k\right)
			\label{SIdef:psi_from_phi}
		\end{equation}
		with $\bm{Z}':= (Z_2,\dots,Z_K)$, we can 
		rewrite Eq.~\eqref{SIeq:asymptoticEq_sfda} as 
		\begin{equation}
			\int_{-\infty}^{\infty}dy\rho(y)\psi\left(W-y;\bm{Z}'\right)-\psi\left(W;\bm{Z}'\right)\approx 0.
		\end{equation}
		This form of the integral equation is useful because the dependence on $y$ disappears for $\bm{Z}'$, and can be regarded as an effectively one-dimensional integral equation.  

		With the condition that the mark distribution has zero mean $m:=\int_{-\infty}^\infty y\rho(y)dy=0$ with fast-decaying tail, the solution of this integral equation is given by
		\begin{equation}
			\psi\left(W;\bm{Z}'\right)=C_0\left(\bm{Z}'\right) + WC_1\left(\bm{Z}'\right)
			\label{eq:sol_integral_eq_const}
		\end{equation}
		with arbitrary functions $C_0(\bm{Z}')$ and $C_1(\bm{Z}')$ that do not have $W$ as an argument (see Appendix~\ref{sec:Der_sol_integral_equation} for the derivation). 
			As confirmed soon later, the natural boundary condition requires $C_1(\bm{Z}')=0$ and thus the general solution is finally given by 
			\begin{equation}
				\psi\left(W;\bm{Z}'\right)=C_0\left(\bm{Z}'\right).
				\label{eq:sol_integral_eq_const2}
			\end{equation}
		The tension distribution in the steady state $P_{\mathrm{ss}}(\nu):= \lim_{t\to \infty}\la \delta(\nu-\nu(t))\ra$ 
		is given by marginalisation of the full distribution as 
		\begin{equation}
			P_{\mathrm{ss}}(\nu) := \int_{-\infty}^\infty d\bm{z}P_{\mathrm{ss}}(\bm{z})\delta \left(\nu-\sum_{k=1}^Kz_k\right) \propto {1\over g(\nu)} \>\>\> \mbox{for large }\nu, 
			\label{SIeq:marginalization_calc}
		\end{equation}
		assuming that $\int_{-\infty}^{\infty}{C_0(z'_2,\dots, z'_K)\Pi_{j=2}^Kdz'_j}$ is finite (see also Appendix~\ref{sec:app:Der_Marginalization} for the detailed calculation below). This implies Eq.~\eqref{eq:steady_PDF} in the main text. 	
		This implies that the steady intensity PDF is given by 
		\begin{equation}
			P_{\mrss}(\lambda) \propto \lambda^{-1}\left\{\frac{dg(\nu)}{d\nu}\right\}^{-1}\bigg|_{\nu = g^{-1}(\lambda)}
		\end{equation}
		where we have used the Jacobian relationship $P_{\mrss}(\lambda) = |d\nu/d\lambda|P_{\mrss}(\nu)$, representing the conservation of probability 
		under a change of variable. 

	\subsubsection{Consistency with the natural boundary condition.}
		Technically, the steady state solution of the master equation should satisfiy the natural boundary condition, requiring a vanishing probability current at $\bm{z}\to \bm{\infty}$. Here we impose this condition on the steady state solution~\eqref{eq:sol_integral_eq_const}. For large $\bm{z}$, the master equation is asymptotically given by 
		\begin{equation}
			\frac{\partial P_t(\bm{z})}{\partial t} \simeq 
			\int_{-\infty}^{\infty}dy\rho(y)\left\{G(\bm{z}-y\bm{\tilh})P_t(\bm{z}-y\bm{\tilh})-G(\bm{z})P_t(\bm{z})\right\},
		\end{equation}
		which is equivalent to 
		\begin{equation}
			\frac{\partial P_t(W;\bm{Z}')}{\partial t} \simeq 
			\int_{-\infty}^{\infty}dy\rho(y)\left\{G(W-y;\bm{Z}')P_t(W-y;\bm{Z}')-G(W;\bm{Z}')P_t(W;\bm{Z}')\right\},
		\end{equation}
		after the variable transformation $\bm{z}\to \bm{Z}:=(W;\bm{Z}')$ defined by Eq.~\eqref{SIeq:var_trans_intEq}. The probability current is defined by the Kramers-Moyal expansion: 
		\begin{equation}
			\frac{\partial P_t(W;\bm{Z}')}{\partial t} \simeq -\frac{\partial}{\partial W}J_t(W;\bm{Z}'), \>\>\> 
			J_t(W;\bm{Z}') := \sum_{n=1}^\infty \frac{(-1)^{n-1} \alpha_{n}}{n!}\frac{\partial^{n-1}}{\partial W^{n-1}}G(W;\bm{Z}')P_t(W;\bm{Z}'), \>\>\>
			\alpha_{n}:=\int_{-\infty}^\infty dy y^n\rho(y). 
		\end{equation}
		The natural boundary condition requires the vanishing probability current at $W=+\infty$ as 
		\begin{equation}
			\lim_{W\to +\infty}J_{\mrss}(W;\bm{Z}') = 0.
		\end{equation}
		Since $\alpha_1=m=0$ for the zero-mean mark $m=0$, in the steady state, the natural boundary condition is given by
		\begin{equation}
			J_{\mrss}(W;\bm{Z}') = \sum_{n=2}^\infty \frac{(-1)^{n-1}\alpha_{n}}{n!}\frac{\partial^{n-1}}{\partial W^{n-1}}\left(C_0(\bm{Z}')+WC_1(\bm{Z}')\right) = -\frac{\alpha_2}{2} C_1(\bm{Z}') = 0,
		\end{equation}
		which requires that $C_1(\bm{Z}')=0$. 

	\subsubsection{Polynomial intensity case: $g(\nu)\propto \nu^n$ for some $n>2$}
		We next study the power law forms of the intensity, by assuming various tension-intensity maps. Let us first consider the polynomial case of $g(\nu) = \lambda_0 + \lambda_1 \nu^n$ for some $n>2$. This means that
		\begin{equation}
			\frac{dg}{d\nu} = n\lambda_1 \nu^{n-1}, \>\>\> g^{-1}(\lambda) = \left(\frac{\lambda-\lambda_0}{\lambda_1}\right)^{1/n} \propto \lambda^{1/n} \>\>\> \mbox{for large }\lambda. 
		\end{equation}
		This means that the asymptotic form is given by the quasi-Zipf law: 
		\begin{equation}
			P_{\mrss}(\lambda) \propto \lambda^{-2+1/n} = \lambda^{-1-a}, \>\>\> a:= 1-\frac{1}{n}.
		\end{equation}

	\subsubsection{Superpolynomial intensity case $g(\nu)>O(\nu^n)$ with any $n>2$}
		To develop some intuition, let us consider two typical cases.  
		One typical case is given by $g(\nu)=\lambda_0 e^{\beta \nu}$, implying that 
		\begin{equation}
			\frac{1}{g(\nu)}\frac{dg}{d\nu} = \beta, \>\>\> 
			g^{-1}(\lambda) = \frac{1}{\beta}\log \frac{\lambda}{\lambda_0}
			\>\>\> \Longrightarrow \>\>\> 
			P_{\mrss}(\lambda) \propto \lambda^{-2}. 
		\end{equation}
		Another typical case is given by the super-exponential case $g(\nu) = \lambda_0 e^{\beta \nu^n}$ with $n > 1$. This case implies
		\begin{equation}
			\frac{1}{g(\nu)}\frac{dg}{d\nu} = n \beta \nu^{n-1}, \>\>\> 
			g^{-1}(\lambda) = \left(\frac{1}{\beta}\log \frac{\lambda}{\lambda_0}\right)^{1/n}, 
			\>\>\> \Longrightarrow \>\>\>
			P_{\mrss}(\lambda) \propto \lambda^{-2} \left(\log \lambda\right)^{-1+1/n}\>\>\> \mbox{for large }\lambda. 
		\end{equation}
		This means that Zipf's law holds up to the minor logarithmic factor. 
		
		Let us generalise these Zipf's law for general superpolynomial cases as follows. Considering the relation
		\begin{equation}
			\frac{1}{g(\nu)}\frac{dg}{d\nu} = \frac{d}{d\nu} \log g(\nu),
		\end{equation}
		we obtain
		\begin{equation}
			 P_{\mrss}(\lambda) \propto \lambda^{-1}\left\{ \frac{dg}{d\nu}\right\}^{-1} 
			 = \lambda^{-1} \frac{1}{g(\nu)} \left\{ \frac{d}{d\nu}\log g(\nu)\right\}^{-1}
			 = \lambda^{-2} \left\{ \frac{d}{d\nu}\log g(\nu)\right\}^{-1}
			 = \lambda^{-2} \left\{ \frac{d}{d\nu}\log \lambda(\nu)\right\}^{-1}.
		\end{equation}
		For most of physically motivated functions $\lambda = g(\nu)$, the logarithmic contribution from $\left\{\log \lambda \right\}^{-1}$ is subleading compared with Zipf's part $\lambda^{-2}$ except for the polynomial intensity. Indeed, by assuming the asymptotic balance between the logarithmic and the power law parts as $(d/d\nu)\log \lambda(\nu) \simeq C\lambda^{a}$ with some real numbers $a\neq 0$ and $C < \infty$, we deduce the polynomial intensity $\lambda (\nu)= g(\nu) \propto \nu^{-1/a}$ as the corresponding exception. Thus, we find that the logarithmic factor $\left\{ (d/d\nu)\log \lambda(\nu)\right\}^{-1}$ is a minor correction term for the superpolynomial cases. 
		
	\subsection{Solution for negative-mean mark distributions}
		The above calculation can be generalised by assuming that the mean mark is negative and that the probability of the positive marks is nonzero: 
		\begin{equation}
			m := \int_{-\infty}^\infty y\rho(y)dy < 0, \>\>\> 
			m_+ := \int_{0}^\infty y\rho(y)dy > 0. 
			\label{SIeq:cond_negative_mean}
		\end{equation}
		For large $z$, the steady master equation can be rewritten as 
		\begin{equation}
			\int_{-\infty}^{\infty}dy\rho(y)\psi\left(W-y;\bm{Z}'\right)-\psi\left(W;\bm{Z}'\right)\approx 0
		\end{equation}
		with $\psi(W;\bm{Z}'):=G(\bm{z})P_{\mrss}(\bm{z})$ after the variable transformaton $\bm{z}\to \bm{Z}:=(W;\bm{Z}')$ defined by Eq.~\eqref{SIeq:var_trans_intEq}. Under the condition~\eqref{SIeq:cond_negative_mean}, the general solution is given by 
		\begin{equation}
			\psi\left(W;\bm{Z}'\right) = C_0(\bm{Z}')e^{-c^*W} + C_1(\bm{Z}'),
			\label{SIeq:sol_master_negativeM}
		\end{equation}
		where $C_0(\bm{Z}')$ and $C_1(\bm{Z}')$ are arbitrary functions without $W$ as an argument (see Appendix~\ref{sec:Der_sol_integral_equation} for the derivation) and $c^*>0$ is the unique positive root of $\Phi(c^*)=0$ for the moment-generating function defined by
		\begin{equation}
			\Phi(x):= \int_{-\infty}^\infty \rho(y)(e^{xy}-1)dy.
		\end{equation}
		We can prove that $\Phi(c)=0$ has only two roots at $c=0$ and $c=c^*>0$ as shown in Appendix~\ref{sec:Der_sol_integral_equation}. The outline of the proof is as follows: since the second order derivative of $\Phi(x)$ is always positive (see Eq.~\eqref{SIeq:CGF_convex} below), the first order derivative is an increasing function of $x$. If the mean of $y$ is zero, at $x=0$, the first order derivative is equal to zero, and thus positive for $x>0$. This proves that $c^*=0$ for $m=0$.  If the mean of $y$ is negative, the first order derivative is negative at $x=0$ but it increases and passes positive for larger $x$. Thus $\Phi(x)$ first decreases below 0 and then crosses it again at some $c^*$, which is the solution. 
		
		Finally, the natural boundary condition requires that $C_1(\bm{Z}')$ must be zero as shown later soon: $C_1(\bm{Z}')=0$. We then obtain the general solution
		\begin{equation}
			\psi\left(W;\bm{Z}'\right) = C_0(\bm{Z}')e^{-c^*W}. 
			\label{SIeq:sol_master_negativeM2}
		\end{equation}
		The tension distribution in the steady state $P_{\mathrm{ss}}(\nu):= \lim_{t\to \infty}\la \delta(\nu-\nu(t))\ra$ is given by marginalisation of the full distribution as 
		\begin{equation}
			P_{\mathrm{ss}}(\nu) := \int_{-\infty}^\infty d\bm{z}P_{\mathrm{ss}}(\bm{z})\delta \left(\nu-\sum_{k=1}^Kz_k\right) \propto {1\over g(\nu)}\exp\left(-\frac{c^*}{\tilh_{\rm tot}}\nu\right) \>\>\> \mbox{for large }\nu
			\label{SIeq:marginalization_calc_negativeM}
		\end{equation}
		with
		\begin{equation}
			\tilh_{\rm tot} := \sum_{k=1}^K\tilh_k = h(t=0),
		\end{equation}
		which is derived in Appendix~\ref{sec:app:Der_Marginalization_negativeM}. This implies that the steady intensity PDF is given by 
		\begin{equation}
			\label{eq:steady_sol_asymmetric_intensity_gen}
			P_{\mrss}(\lambda) \propto \lambda^{-1}\left[\exp\left(-\frac{c^*}{h(0)}\nu\right)\left\{\frac{dg(\nu)}{d\nu}\right\}^{-1}\right]_{\nu = g^{-1}(\lambda)}
		\end{equation}
		where we have used the Jacobian relationship $P_{\mrss}(\lambda) = |d\nu/d\lambda|P_{\mrss}(\nu)$, representating the probability conservation.

		\subsubsection{Consistency with the natural boundary condition.}
			Let us confirm the consistency of the general solution~\eqref{SIeq:sol_master_negativeM} for the natural boundary condition. By substituting the general solution~\eqref{SIeq:sol_master_negativeM} in the probability current $J_{\mrss}(W;\bm{Z}')$, we obtain 
			\begin{align}
				J_{\mrss}(W;\bm{Z}') = \sum_{n=1}^\infty \frac{(-1)^{n-1}\alpha_n}{n!}\frac{\partial^{n-1}}{\partial W^{n-1}}\left(C_1(\bm{Z}')+e^{-c^*W}C_0(\bm{Z}')\right)
				= m C_1(\bm{Z}') + \frac{C_0(\bm{Z}')}{c^*}\Phi(c^*)e^{-c^*W},
				\label{SIeq:KM_negativeM}
			\end{align}
			where we have used the expansion of the generating function $\Phi(x)=\sum_{n=1}^\infty (\alpha_n/n!)x^n$. Since $\Phi(c^*)=0$, we obtain $J_{\mrss}(W;\bm{Z}') = m C_1(\bm{Z}')$. Because the natural boundary condition requires $\lim_{W\to \infty}J_{\mrss}(W;\bm{Z}') = 0$ for any $\bm{Z}'$, the function $C_1(\bm{Z}')$ must be zero. 

		\subsubsection{Exponential intensity case: $g(\nu)\simeq \lambda_0e^{\beta \nu}$}
			For simplicity, let us consider the exponential intensity 
			\begin{equation}
				\lambda = g(\nu) = \lambda_0 e^{\beta \nu}, \>\>\> \beta >0. 
			\end{equation}
			Using formula~\eqref{eq:steady_sol_asymmetric_intensity_gen}, the steady state PDF is given by the power law intensity distribution
			\begin{equation}
				P_{\mathrm{ss}}(\lambda) \propto \lambda^{-2 - \beta^{-1}a}, \>\>\> a:= \frac{c^*}{h(0)},
			\end{equation}
			suggesting a thinner tail for the negative-mean mark distributions $m<0$. 

		\subsubsection{Super-exponential intensity case: $g(\nu)\simeq \lambda_0e^{\beta \nu^n}$ with $n>1$}
			We next consider the super-exponential intensity 
			\begin{equation}
				\lambda = g(\nu) = \lambda_0 e^{\beta \nu^n}, \>\>\> \beta >0, \>\>\> n>1. 
			\end{equation}
			Using formula~\eqref{eq:steady_sol_asymmetric_intensity_gen}, the steady PDF is given by 
			\begin{align}
				P_{\mathrm{ss}}(\lambda) &\propto \lambda^{-2}\left(\log \frac{\lambda}{\lambda_0}\right)^{-1+1/n}\exp\left(-\frac{c^*}{h(0)}\left(\frac{1}{\beta}\log\frac{\lambda}{\lambda}\right)^{1/n}\right) \notag \\
				&= \exp\left[-2\log \lambda -\frac{c^*}{h(0)}\left(\frac{1}{\beta}\log\frac{\lambda}{\lambda_0}\right)^{1/n} + \left(-1+\frac{1}{n}\right)\log \log \frac{\lambda}{\lambda_0}\right] \>\>\> \mbox{ for large }\lambda.
				\label{SIeq:asymptotic_PDF_superexpon_asymmetric}
			\end{align}
			Here we can drop the sub-dominant double-logarithmic correction, since
			\begin{equation}
				\lim_{\lambda \to \infty}\frac{1}{\log \lambda}\log \log \frac{\lambda}{\lambda_0} = 0.
			\end{equation}
			We thus obtain the asymptotic formula for the super-exponential cases to leading order: 
			\begin{equation}
				P_{\mrss}(\lambda) \propto \lambda^{-2-\beta^{-1}a \left(\beta^{-1}\log \frac{\lambda}{\lambda_0}\right)^{-1+1/n}} \>\>\> \mbox{ for large }\lambda, \>\>\> a:= \frac{c^*}{h(0)},
			\end{equation}
			which obeys quasi-Zipf's scaling with the correction in the exponent due to the asymmetry of the mark distribution. 
			
			One can notice that the correction term $(\log \lambda /\lambda_0)^{1/n}$ in Eq.~\eqref{SIeq:asymptotic_PDF_superexpon_asymmetric} can be also regarded as subleading in the sense that $\lim_{\lambda \to \infty}\frac{1}{\log \lambda} \left(\beta^{-1}\log \frac{\lambda}{\lambda_0}\right)^{1/n} = 0$, deducing the Zipf's law $P_{\mrss}(\lambda) \propto \lambda^{-2}$ for large $\lambda$. However, one should be cautious in dropping this term for practical analyses, since the convergence speed is slow.

		\subsubsection{Polynomial intensity case: $g(\nu)\simeq \lambda_0\nu^n$ with $n>2$}
			We also consider the polynomial intensity 
			\begin{equation}
				\lambda = g(\nu) = \lambda_0\nu^n, \>\>\> n>2. 
			\end{equation}
			Using formula~\eqref{eq:steady_sol_asymmetric_intensity_gen}, the steady state PDF is given by the power law intensity distribution with the streched-exponential truncation: 
			\begin{equation}
				P_{\mathrm{ss}}(\lambda) \propto \lambda^{-2+1/n}\exp\left[-a\left(\frac{\lambda}{\lambda_0}\right)^{1/n}\right] \>\>\> \mbox{ for large }\lambda, \>\>\> a:= \frac{c^*}{h(0)}.
			\end{equation}
	
\subsection{Markov embedding and field master equation for continuous sum of exponentials}\label{sec:SI_continuousSum}
	We have formulated the Markov embedding method for the nonliear Hawkes process with the discrete sum of exponentials~\eqref{SIeq:superpositionOfExpon} 
	and have derived the corresponding master equation~\eqref{SIeq:master_discrete}. 
	Here we formally generalise this methodology for the most general case of continuous sum of exponentials:
	\begin{equation}
		h(t)=\int_{0}^{\infty}dx\tilde{h}(x)e^{-t/x}. 
	\end{equation}
	On the basis of this decomposition, the original nonlinear Hawkes process is converted into a Markovian stochastic partial differential equation (SPDE).
	\begin{equation}
		\nu(t) = \int_0^\infty dx z(t,x), \>\>\> 
		{\partial z(t,x) \over \partial t} = -{z(t,x)\over x} + \tilh(x) \xi^{\mathrm{P}}_{\rho(y); \lambda(t)}.
		\label{SIeq:MarkovSPDE}
	\end{equation}
	This conversion implies that the original one-dimensional non-Markovian process $\nu(t)$ is equivalent to an infinite-dimensional Markovian process described by $\{z(t,x)\}_{x\in (0,\infty)}$. 
	Here $x \in (0,\infty)$ is the label of the auxiliary variables $\{z(t,x)\}_{x\in (0,\infty)}$ distributed on the auxiliary field $(0,\infty)$. 
	
	The master equation corresponding to the SPDE~\eqref{SIeq:MarkovSPDE} can be formally written 
	with the formalism of functional calculus. 
	Indeed, by introducing the probability density functional $P[z]:=P_t[\{z(x)\}_x]$ and the intensity functional $G[z]:=g\left(\int_0^\infty dx z(t,x)\right)$,
	the field master equation~\cite{KS_PRL2020,KS_PRR2020} is given by
	\begin{equation}
		\frac{\partial P_t[z]}{\partial t}=\int_{0}^\infty dx \frac{\delta}{\delta z(x)}\frac{z(x)}{x}P_t[z]+\int_{-\infty}^{\infty}dy\rho(y)\left\{G[z-y\tilde{h}]P_t[z-y\tilde{h}]-G[z]P_t[z]\right\},
		\label{SIeq:master_continuous}
	\end{equation}
	which should be interpreted as a formal limit from the discrete representation~\eqref{SIeq:master_discrete} according to the standard convention~\cite{GardinerB},
	and thus has the same asymptotic solution~\eqref{SIeq:marginalization_calc}.

	The functional description for the field master equation~\eqref{SIeq:master_continuous} is formally introduced as follows. 
	Let us discuss the nonlinear Hawkes process~\eqref{SIeq:SDE_Markov_discrete} for the discrete sum of exponentials~\eqref{SIeq:superpositionOfExpon}, whose master equation is given by Eq.~\eqref{SIeq:master_discrete}. 
	Here we introduce a lattice for $x$ with interval $dx$, such that
	\begin{equation}
		\tau_k = x_k = kdx, \>\>\> \tilh(x_k)dx = \tilh_k, \>\>\> z_k(t) = z(t,x_k)dx
	\end{equation}
	for the non-negative integer $k=1,\dots, K$. 
	Equation~\eqref{SIeq:SDE_Markov_discrete} is then rewritten as
	\begin{equation}
		\nu(t) = \sum_{k=1}^K z(t,x_k), \>\>\> 
		\frac{\partial z(t,x_k)}{\partial t} = - \frac{z(t,x_k)}{x_k} + \tilh_k \xi^{\rm P}_{\rho(y);\lambda(t)}, \>\>\> 
		h(t) = \int_0^\infty dx\tilh(x)e^{-t/x} \approx \sum_{k} \tilh_ke^{-t/x}.
	\end{equation}
	The corresponding master equation is given by
	\begin{equation}
				\frac{\partial P_t(\bm{z})}{\partial t}=\sum_{k=1}^{K} dx\frac{1}{dx}\frac{\partial}{\partial z(x_k)}\frac{z(x_k)}{x_k}P_t(\bm{z})
		+\int_{-\infty}^{\infty}dy\rho(y)\left\{G(\bm{z}-y\bm{\tilh})P_t(\bm{z}-y\bm{\tilh})-G(\bm{z})P_t(\bm{z})\right\},
	\end{equation}
	by introducing a vector $\bm{\tilh}:=(\tilh_1,\dots,\tilh_K)$. Let us take the formal limit $K\to \infty$ and $dx \downarrow 0$ to deduce the field master equation~\eqref{SIeq:master_continuous} by replacement
	\begin{equation}
		\frac{\delta }{\delta z(x)} [...]:= \lim_{dx\downarrow 0}\lim_{K\to \infty}\frac{1}{dx}\frac{\partial}{\partial z(x_k)} [...], \>\>\> \int_0^\infty dx [...] = \lim_{dx\downarrow 0}\lim_{K\to \infty}\sum_{k=1}^K dx[...],
	\end{equation}
	which follows the convention~\cite{GardinerB}. We note that the rigorous foundation for the functional description has not been established yet~\cite{GardinerB}, and constitutes a problem out of scope of our paper. 

	\section{Illustrative case}
		\label{sec:SM:illustrativeCase}
		As an appendix, let us focus on the illustrative case where the memory function $h(t)=\tilh  e^{-t/\tau}$ is a single exponential and the distribution of marks is symmetric: $\rho(y)=\rho(-y)$, in order to provide an intuitive understanding of the underlying generating mechanism of the Zipf and quasi-Zipf laws.	In this case, the original model can be converted into a simple process obeying the stochastic differential equation (SDE)
		\begin{equation}
			\frac{d\nu}{dt} = -\frac{\nu}{\tau} + \tilh \xi^{\rm P}_{\rho(y); \lambda}
			\label{wyn2trbgq}
		\end{equation}
		in terms of the compound Poisson process $\xi^{\rm P}_{\rho(y); \lambda}$ with jump-size distribution $\rho(y)$ and corresponding intensity $\lambda=g(\nu)$. We apply the diffusive approximation: $\xi^{\rm P}_{\rho(y); \lambda(t)} \approx \sqrt{2Dg(\nu)}\xi^{\rm G}$, with the standard white Gaussian noise $\xi^{\rm G}$ and $D:=(\tilh^2/2) \int_{-\infty}^{\infty} y^2\rho(y)dy$, which requires that the second-order moment of the PDF $\rho(y)$ exists. This can be shown by the Kramers-Moyal (KM) expantion and truncating its series up to the second order. Indeed, the KM expantion of the master equation is given by
		\begin{subequations}
		\begin{align}
			\frac{\partial P_t(\nu)}{\partial t} &= \frac{1}{\tau}\frac{\partial}{\partial \nu}\nu P_t(\nu) + \int_{-\infty}^\infty dy \rho(y)[g(\nu-\tilh y)P_t(\nu-\tilh y) - g(\nu)P_t(\nu)] \notag\\
			&= \frac{1}{\tau}\frac{\partial}{\partial \nu}\nu P_t(\nu) + \sum_{k=1}^{\infty} \frac{\tilh^{2k}\alpha_{2k}}{(2k)!}\frac{\partial^{2k}}{\partial \nu^{2k}} [g(\nu)P_t(\nu)]
			\label{eq:SM:KM_expansion_1}
		\end{align}
		with the $k$the-order KM coefficient defined by
		\begin{equation}
			\alpha_k := \int_{-\infty}^\infty y^k\rho(y)dy.
			\label{eq:SM:KM_expansion_2}
		\end{equation}
		By truncating the KM expansion up to the second-order, we obtain
		\begin{equation}
			\frac{\partial P_t(\nu)}{\partial t} \approx \frac{1}{\tau}\frac{\partial}{\partial \nu}\nu P_t(\nu) + D\frac{\partial^2}{\partial \nu^2}g(\nu)P_t(\nu),
			\label{eq:SM:diffusive_singleExpoen}
		\end{equation}
		\end{subequations}
		which is equivalent to Eq.~\eqref{wyn2trbgq}. 
		Since $\sqrt{2Dg(\nu)} \gg \nu$ for fast-accelerating intensities, we obtain a $\nu$-dependent diffusion process
		\begin{equation}
			\label{eq:diffusion_model}
			\frac{d\nu}{dt} \approx -\frac{\nu}{\tau} + \sqrt{2Dg(\nu)}\xi^{\rm G} \>\>\>\>\> (\mbox{for large }\nu)~.
		\end{equation}
		We note that the truncation of the KM expansion can be proved by making assumption of the diffusive scaling using the system-size expansion (see Appendix~\ref{sec:SM:DiffAsymptotics} for the mathematical detail). 

		\subsection{Assuming the fast-acceralating intensities}
			Let us consider the case of fast-accelerating intensities $g(\nu) > O(\nu^2)$. For this case, the linear part $- \nu/\tau$ becomes smaller than the diffusive term $\sqrt{2D g(\nu)}\xi^{\mathrm{G}}$ at large $\nu$: 
			\begin{equation}
				\sqrt{2D g(\nu)}\xi^{\mathrm{G}} > O(\nu) = O(-\nu/\tau). 
			\end{equation}
			This means that the diffusive model~\eqref{eq:diffusion_model} can be regarded as the inhomogeneous diffusive process for large $\nu$ without relaxation term:
			\begin{equation}
				\frac{d\nu}{dt} \approx \sqrt{2Dg(\nu)}\xi^{\rm G} \>\>\>\>\> (\mbox{for large }\nu)~,
			\end{equation}
			which corresponds to 
			\begin{equation}
				\frac{\partial P_t(\nu)}{\partial t} \approx D\frac{\partial^2}{\partial \nu^2}g(\nu)P_t(\nu) 
				\>\>\>\>\> (\mbox{for large }\nu)~.
			\end{equation}
			This means that the corresponding steady state PDF is asymptotically given by
			\begin{equation}
				\label{eq:steady_PDF}
				P_{\rm ss}(\nu) \approx \frac{1}{g(\nu)} = o(\nu^{-2}) \>\>\>\>\> (\mbox{for large }\nu),
			\end{equation}
			which is consistent with the normalisation condition $\int_{-\infty}^{\infty }d\nu P_{\mrss}(\nu) = 1$ under the assumption of fast-accelerating intensities $g(\nu) > O(\nu^2)$. Since this asymptotic solution is consistent with the normalisation condition, the solution~\eqref{eq:steady_PDF} is the correct asymptotic form for $P_{\rm ss}(\nu)$. Since $\lambda=g(\nu)$, the identity $P_{\mrss}(\nu) d\nu = P_{\mrss}(\lambda)d\lambda$ expressing the conservation of probability under a change of variable leads to the following expression for the PDF of $\lambda$:
			\begin{equation}
				P_{\mrss}(\lambda) \propto \frac{1}{\lambda}
				\left\{ \frac{dg}{d\nu}\left( g^{-1}(\lambda) \right)\right\}^{-1}.
			\end{equation}
			This recovers the universal and quasi-Zipf's laws~\eqref{eq:Zipf_intensity} for both superpolynomial and polynomial fast-accelerating intensity as shown in Appendix~\ref{SIsec:sol_zero-mean}. 
				
		\subsubsection{Intuitive understanding}			
			An intuitive understanding that the SDE~\eqref{eq:diffusion_model} leads to a stationary solution and thus a bona fide PDF for $\nu$ is obtained by applying Ito's lemma on the change of variable  $\nu \to \chi := e^{-(\beta /2) \nu}$ for the case~\eqref{eq:ExpIntensity} with $D=1/2$ and $\lambda_0=1$, leading to
			\begin{equation}
				\label{wj4i7k4uj3}
				d\chi \approx \mu_\chi dt  + {\beta \over 2} dB~, ~~{\rm with}~ 
				\mu_\chi : = {\beta^2 \over 8 \chi} - {\chi \ln \chi \over \tau},
			\end{equation}
			where we use the mathematical notation $dB := \xi^{\rm G} dt$ to represent the increment of the standard Brownian (or Wiener) process.
			Expression~\eqref{wj4i7k4uj3} describes the motion of a Brownian particle in the potential 
			$V(\chi) = - \int^{\chi} \mu_{\chi'} d\chi' = ({1 / 2\tau}) \chi^2 \ln ( \chi e^{-2}) - ({\beta^2 / 8}) \ln \chi$, from which the drift force $\mu_{\chi}$ derives. The behaviour of $\lambda$ at large values is controlled by the dynamics of $\nu$ at large positive values, 
			which corresponds to $\chi$ close to $0$.  As $\chi$ approaches $0$, $\mu_{\chi}$ diverges on the positive side and repeals $\chi$ from the origin. Thus $\nu$ and $\lambda$ never diverges. 
			When $\chi$ grows, $\mu_{\chi}$ becomes negative and also diverge in amplitude, pushing it back to smaller values, thus preventing $\nu$ to become too negative and therefore stopping $\lambda$ from being too small. 

		\subsection{The case of the quadratic Hawkes process}
			\label{sec:SM:QHawkes}
			As a marginal case, let us consider the case with a quadratic intensity
			\begin{equation}
				g(\nu) = \lambda_0 + \lambda_1 \nu^2 = O(\nu^2),
			\end{equation}
			which does not belong to the fast-accelerating intensity and is out of scope of our main manuscript. A part of this case is studied in Ref.~\cite{QHawkesBouchaud}, and this non-linear Hawkes process is called the quadratic Hawkes (QHawkes) processes exhibiting quite different behaviour (e.g., interested readers should see Ref.~\cite{QHawkesBouchaud}, where a special case of the QHawkes process is investigated by assuming the diffusive limit and the exponential memory kernel $h(t)=\tilh e^{-t/\tau}$. Before our work, this was the only study where an explicit asymptotic solution of the nonlinear Hawkes processes was given for a specific setup). 

			The QHawkes process belongs to another class of nonlinear Hawkes processes, because the relaxation term $-\nu/\tau$ and the diffusive term $\sqrt{2Dg(\nu)}\xi^{\rm G} \propto \sqrt{2\lambda_1 D}\nu \xi^{\rm G}$ are of the same order for large $\nu$: 
			\begin{equation}
				O(-\nu/\tau) = O\left( \sqrt{2Dg(\nu)}\xi^{\rm G} \right) = O(\nu).
			\end{equation}
			Indeed, the QHawkes is essentially similar to the Kesten process~\cite{Kesten1973} because the diffusive model~\eqref{eq:diffusion_model} can be asymptotically regarded as a continuous version of the Kesten process as 
			\begin{equation}
				\nu(t+dt) \approx \left\{1 -\frac{dt}{\tau} + \sqrt{2\lambda_1 D}dB(t) \right\}\nu(t)
				\>\>\>\>\> (\mbox{for large }\nu)
			\end{equation}
			with the increment of the standard Brownian process $dB:=\xi^{\mathrm{G}}dt$. 
			Since the solution of the Kesten processes are known to obey non-universal power laws (in the sense that the power law exponents continuously vary according to system parameters), the intensity distribution of the QHawkes process also obeys a power law relation, 
			\begin{equation}
				P_{\mrss}(\lambda) \propto \lambda^{-3/2-a}
			\end{equation}
			with a non-universal positive number $a$ which can take any positive number according to system parameters, such as $D$, $\tau$ and $\lambda_1$. We note that this result can be confirmed by directly solving the Fokker-Planck equation~\eqref{eq:SM:diffusive_singleExpoen}, even without knowing the theoretical background of the linear Kesten processes. This is in contrast with the universal Zipf's law in our work under the symmetric assumption $\rho(y)=\rho(-y)$ with a fixed power law exponent independent of system parameters. Thus, the QHawkes process is essentially different from our nonlinear Hawkes models because the QHawkes process can be regarded as a linear-Kesten family member while our setup belongs to nonlinear Kesten families. 

\section{Diffusive asymptotics: the system-size expansion}
	Here we briefly review the system-size expansion, an established perturbative method for systems with small-noise or 
	with weak-coupling. This method is relevant to the diffusive approximation in Appendix~\ref{sec:SM:illustrativeCase} and Zipf's law~\eqref{eq:Zipf_eventNumbers} for the number of events in the main text.  

	In Appendix~\ref{sec:SM:illustrativeCase}, we have used the diffusive approximation to derive multiplicative Gaussian noise terms $\sqrt{\lambda}\xi^{\mathrm{G}}$ from the Poisson noise terms $\xi^{\mathrm{P}}_{\rho(y);\lambda}$. This approximation is asymptotically correct by assuming the diffusive scaling
	\begin{equation}
		g(\nu) = \frac{1}{\eps^2}\bar{g}(\nu), \>\>\> 
		\rho(y) = \frac{1}{\eps}\bar{\rho}\left(\frac{y}{\eps}\right)
		\label{eq:SM:scaling_diffusive}
	\end{equation}
	with a small parameter $\eps > 0$. We also assume that $\bar{g}(\nu)$ and $\bar{\rho}(Y)$ are $\eps$-independent with scaled mark $Y:=y/\eps$. We note that this method is essentially equivalent to the system-size expansion (or the $\Omega$ expansion), invented by van Kampen~\cite{VanKampen}; historically $\eps$ is often written by $\eps:= 1/\Omega$ with large parameter $\Omega$, called the system size (see a recent related Letter~\cite{KzPRL2015} and a review~\cite{KzBook} for more detail). 

	The intuitive explanation of this scaling~\eqref{eq:SM:scaling_diffusive} is given as follows: the compound Poisson noise is given by $\xi^{\mathrm{P}}_{\rho(y);g(\nu)} = \sum_{i=1}^{N(t)}y_i \delta (t-t_i)$. Here we assume that the mark $y_i$ is sufficiently small; $y_i$ is proportional to a small parameter $\eps$ as 
	\begin{equation}
		y_i:=\eps Y_i, \>\>\> \xi^{\mathrm{P}}_{\rho(y);g(\nu)} = \sum_{i=1}^{N(t)}\eps Y_i \delta (t-t_i) = \eps \xi^{\mathrm{P}}_{\bar{\rho}(y);g(\nu)}
	\end{equation}
	with $\eps$-independent mark $Y_i$ and mark distribution $\bar{\rho}(y)$. Considering the Jacobian relation associated with the preservation of probability, we obtain
	\begin{equation}
		\rho(y)dy = \bar{\rho}(Y)dY \>\>\> \Longleftrightarrow \>\>\>
		\rho(y) = \frac{1}{\eps}\bar{\rho}\left(\frac{y}{\eps}\right).
		\label{eq:SM:scaling_diffusive2}
	\end{equation}
	In this sense, the scaling~\eqref{eq:SM:scaling_diffusive2} can be regarded as a small-noise limit or a weak-interaction limit. However, if we naively take the small noise limit $\eps\to 0$, the effect of the noise completely disappears. To keep the minimal effect of the noise, let us assume that the intensity is sufficiently large as 
	\begin{equation}
		g(\nu) = \frac{1}{\eps^2}\bar{g}(\nu). 
	\end{equation}
	We thus obtain the diffusive scaling~\eqref{eq:SM:scaling_diffusive}. In this sense, this scaling implies that the mark size is small but the frequency  $\sim \eps^{-2}$ is sufficiently high. We thus obtain the following specific form of the nonlinear Hawkes process (see Fig.~\ref{fig:SM:DiffTrajectory} for typical trajectories):
	\begin{equation}
		\lambda (t) = \frac{1}{\eps^2}\bar{g}\left(\eps\sum_{i=1}^{N(t)} Y_i h(t-t_i)\right).
		\label{eq:SM:point_process_diffusiveScaling}
	\end{equation}
	\label{sec:SM:DiffAsymptotics}
	\begin{figure*}
		\centering
		\includegraphics[width=150mm]{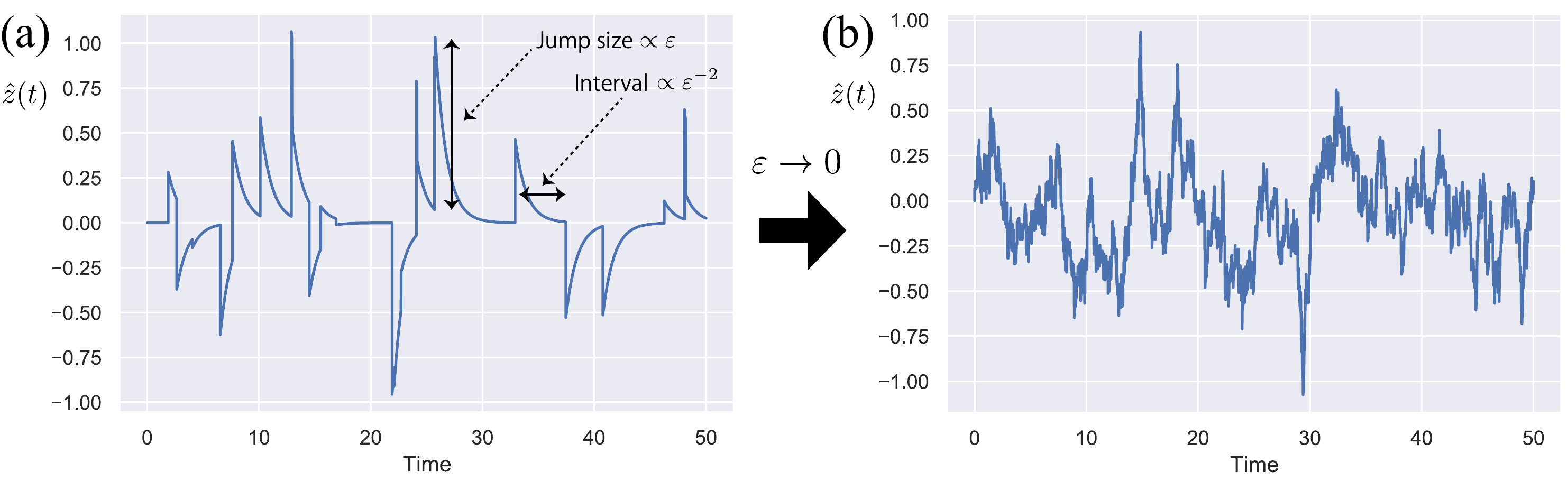}
		\caption{
			Typical trajectories of the nonlinear Hawkes process for the diffusive limit~\eqref{eq:SM:scaling_diffusive2}. Typically, the inter-events interval is proportional to $\eps^{2}$ and the jump-size is proportional to $\eps$. The original point process~\eqref{eq:SM:point_process_diffusiveScaling} gradually reduces to the Langevin dynamics~\eqref{eq:SM:multiplicativeLangevin} for small $\eps$. 
		}
		\label{fig:SM:DiffTrajectory}
	\end{figure*}

	Under this assumption, we can prove that the KM expansion~\eqref{eq:SM:KM_expansion_1} converges to the Fokker-Planck equation~\eqref{eq:SM:diffusive_singleExpoen}. Indeed, the KM coefficients~\eqref{eq:SM:KM_expansion_2} have the scalings
	\begin{equation}
		\alpha_{k} = \int_{-\infty}^\infty y^k\rho(y)dy = 
		\eps^k\int_{-\infty}^\infty Y^k\bar{\rho}(Y)dY = \eps^k \bar{\alpha}_{k}, \>\>\>
		\bar{\alpha}_{k} := \int_{-\infty}^\infty Y^k\bar{\rho}(Y)dY. 
	\end{equation}
	The KM expansion~\eqref{eq:SM:KM_expansion_1} therefore can be transformed as 
	\begin{align}
		\frac{\partial P_t(\nu)}{\partial t} 
		&= \frac{1}{\tau}\frac{\partial}{\partial \nu}\nu P_t(\nu) + \sum_{k=1}^{\infty} \eps^{2(k-1)}\frac{\tilh^{2k} \bar{\alpha}_{2k}}{(2k)!}\frac{\partial^{2k}}{\partial \nu^{2k}} [\bar{g}(\nu)P_t(\nu)] \notag \\
		&= \frac{1}{\tau}\frac{\partial}{\partial \nu}\nu P_t(\nu) +  \frac{\tilh^{2} \bar{\alpha}_{2}}{2}\frac{\partial^{2}}{\partial \nu^{2}} \bar{g}(\nu)P_t(\nu) + O(\eps^2).
	\end{align}
	We thus asymptotically obtain the Fokker-Planck equation~\eqref{eq:SM:diffusive_singleExpoen} for the diffusive scaling. In addition, this Fokker-Planck equation is equivalent to a multiplicative Langevin dynamics described by 
	\begin{equation}
		\frac{d\nu}{dt} = -\frac{\nu}{\tau} + \sqrt{2D \bar{g}(\nu)}\xi^{\mathrm{G}}, \>\>\> D:= \frac{\tilh^{2} \bar{\alpha}_{2}}{2}
		\label{eq:SM:multiplicativeLangevin}
	\end{equation}
	by using the Ito convention for the small $\eps$ limit. This methodology can be readily generalised for general memory kernel $h(t)$, by considering the system-size expansion for the field-master equation.

\section{Zipf's law for the events-number statistics}
	We have studied Zipf's law for the steady distribution of intensity, $P_{\mrss}(\lambda)\propto \lambda^{-1-a}$ by assuming the symmetry $\rho(y)=\rho(-y)$. Since the intensity is one of the fundamental characteristic quantities for point processes in general, our finding will be useful for understanding various Zipf's law even for other quantities. Here we discuss its application to the number of events occurring during a finite time window $\twin$ as an example. 

	Let us consider a long-time interval $[0,T)$ and randomly select a time point $t^*\in [0,T)$. We then count the number of events during an interval $[t^*,t^*+\twin)$ as $N_{\twin}(t^*)$ to observe the corresponding PDF $P_{t=t^*}(N_{\twin})$. In the steady state, we can assume that $P_{t=t^*}(N_{\twin})$ does not depend on the selection of $t^*$ and therefore we write $P_{t=t^*}(N_{\twin})$ by $P_{\mrss}(N_{\twin})$. 

	Here we derive Zipf's law for $P_{\mrss}(N_{\twin})$ using Zipf's law for $P_{\mrss}(\lambda)$, by focusing on the exponential tension-intensity map
	\begin{equation}
		\lambda = g(\nu) = \lambda_0 e^{\beta \nu}.	
	\end{equation}
	The basic idea of the derivation is to use the superposition of the Poisson distributions for a sufficiently short time window $\twin$.
	For a short time window $\twin$, let us assume that the intensity is approximately constant during $[t^*,t^*+\twin)$: $\lambda(t)\approx \mbox{const.}$ for $t\in [t^*,t^*+\twin)$. Under this assumption, the number of events obeys the Poisson distribution:
	\begin{equation}
		P(N_{\twin} | \lambda) \approx \frac{(\lambda \twin)^{N_{\twin}}}{N_{\twin}!}e^{-\lambda \twin}. 
	\end{equation}
	Since $t^*$ is randomly selected, the PDF of $\lambda(t^*)$ obeys Zipf's law~\eqref{eq:Zipf_intensity} for $P_{\mrss}(\lambda)\propto \lambda^{-2}$. We thus derive Zipf's law for the unconditional PDF of the number of events as
	\begin{equation}
		P_{\mrss}(N_{\twin}) = \int_{0}^\infty P(N_{\twin} | \lambda)P_{\mrss}(\lambda)d\lambda \propto N_{\twin}^{-2}. 
		\label{eq:SM:superpositionOfPoisson}
	\end{equation}
	This equation assumes that one can neglect the dependence between the realisations of $\lambda$'s in subsequent windows.  
	This assumption is likely incorrect for large realisations of $\lambda$, giving a clue as to the origin of the deviation
	from Zipf's law beyond the cut-off value $N_{\mathrm{cut}}=O(\eps^{-2})$. The study of this strong dependence regime
	is left for a future work.
	
	While the idea of the superposition of the Poisson statistics works formally, the criteria for the sufficiently-short time window is ambigious for the case of nonlinear Hawkes processes, because the intensity obeys the scale-free distribution (i.e., Zipf's law) and has no clear characteristic timescales as the result of the intemittent properties of the nonlinear Hawkes processes. Because of this scale-free nature, the convergenve of the superposition technique~\eqref{eq:SM:superpositionOfPoisson} is not uniform in terms of $N_{\twin}$ and has a finite cutoff $N_{\mathrm{cut}}$. 

	To clarify this technical issue, let us consider the dimensional analysis of the nonlinear Hawkes processes by assuming the diffusive scaling limit~\eqref{eq:SM:scaling_diffusive}. Under this condition, the Markov-embedding representation~\eqref{SIeq:SDE_Markov_discrete} of the nonlinear Hawkes process is approximated by
	\begin{equation}
		\frac{dz_k}{dt} = -\frac{z_k}{\tau_k} + \tilh_k \xi^{\mathrm{P}}_{\rho(y);\lambda(t)} 
		\approx -\frac{z_k}{\tau_k} + \sqrt{2\bar{D}_k \bar{g}(\nu)} \xi^{\mathrm{G}}, \>\>\> 
		\bar{D}_k:= \frac{1}{2}\tilh_k^2\tilde{\alpha}_2
	\end{equation}
	for small $\eps$. Assuming that $\nu$ is approximately constant $\nu\approx \nu^*:=\nu(t^*)$ during $[t^*,t^*+\twin)$, the typical displacement $\Dzdiff$ due to diffusion is given by
	\begin{equation}
		\Dzdiff^2 \approx 2\bar{D}_k\bar{g}(\nu^*)\twin.
	\end{equation}
	Let us estimate the typical value of $z$. Since the steady PDF of tensions $\nu$ is given by $P_{\mrss}(\nu)\sim e^{-\beta \nu}$, the typical value of $\nu$ is given by $\beta^{-1}$. Since $\nu$ is directly related to $z_k$ as $\nu=\sum_{k=1}^K \nu_k$, the typical value of $z_k$ is also the order of $\beta^{-1}$: $z_k^*=\beta^{-1}$. We thus estimate that the time window $\twin$ is sufficiently short if the following condition is satisfied:
	\begin{equation}
		\Dzdiff \ll z_k^*  \>\>\> \Longleftrightarrow \twin \ll \frac{1}{2\beta^{2}\bar{D}_k\bar{g}(\nu^*)} = \frac{1}{\eps^2}\frac{1}{2\beta^{2}\bar{D}_k\lambda(t^*)}.
	\end{equation}
	Remarkably, $\lambda(t^*)$ obeys a scale-free distribution $P_{\mrss}(\lambda)\propto \lambda^{-2}$ and has no specific characteristic value. It means that this criteria has an explicit dependence on the value of $\lambda(t^*)$. Mathematically, this means that the short time window approximation does not uniformly hold in terms of $\lambda$ if the time window is fixed. Since the number of events during $\twin$ is estimated to be $N \approx \lambda \twin$, the cutoff of the PDF derived from the superposition relation~\eqref{eq:SM:superpositionOfPoisson} is estimated to be 
	\begin{equation}
		N_{\mathrm{cut}} := \lambda(t^*) \twin = \frac{1}{\eps^2}\frac{1}{2\beta^{2}\bar{D}_k} 
		= O(\eps^{-2}).
	\end{equation}
	In this sense, Zipf's law for $P_{\mrss}(N_{\twin})$ holds only up to the cutoff $N_{\twin}\ll N_{\mathrm{cut}}$ as 
	\begin{equation}
		P_{\mrss}(N_{\twin}) \propto N_{\twin}^{-2} \>\>\> (N_{\twin} \ll N_{\mathrm{cut}}). 
	\end{equation}
	Since the cutoff diverges as $\lim_{\eps\downarrow 0}N_{\mathrm{cut}} =\infty$, this asymptotic relation holds for a wide regime for small $\eps$ and should be regarded as an intermediate asymptotics~\cite{BarenblattB} for the diffusive limit (see Fig.~\ref{fig:SM:ZipfForNumberOfEvents} for the numerical confirmation). 
	\label{sec:SM:ZipfForNumberOfEvents}
	\begin{figure*}
		\centering
		\includegraphics[width=180mm]{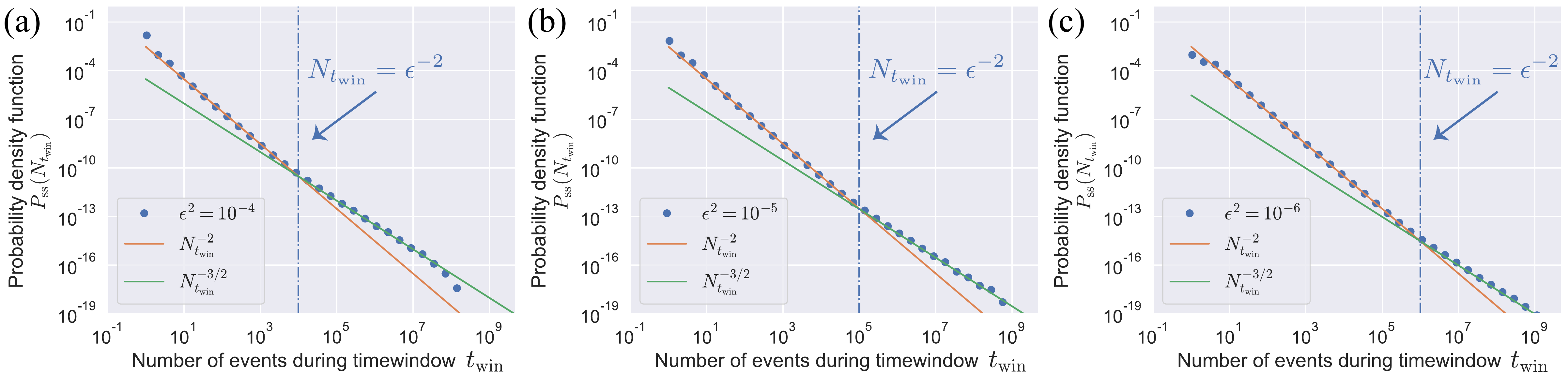}
		\caption{
				Numerical confirmation of Zipf's law for the number of events $N_{\twin}$ during a short time $\twin$. By setting $\twin = 10^{-5}$, we numerically observe the PDF of $N_{\twin}$ for $\eps^2=10^{-4},10^{-5},10^{-6}$. Zipf's law $P_{\mrss}(N_{\twin}) \propto N_{\twin}^{-2}$ was found to hold up to $N_{\mathrm{cut}}=O(\eps^{-2})$ as theoretically predicted, supporting that Zipf's law is indeed valid as intermediate asymptotics. Beyond the cutoff $N>N_{\mathrm{cut}}$, we numerically found a fatter tail characterised by a power law exponent $3/2$. This finding is an interesting issue, requiring further investigation. 
		}
		\label{fig:SM:ZipfForNumberOfEvents}
	\end{figure*}  
	
	\subsubsection*{Numerical scheme}
		For the numerical simulation of Figure~\ref{fig:MarkovEmbedding}c, we set $(\lambda_0,\beta, \sigma, \lambda_{\max}, \twin, \Delta t_{\max}^{(1)}, \Delta t_{\max}^{(2)})=(\eps^{-2},5,\eps,\infty,10^{-5},10^{-6}, 10^{-2})$ for $g(\lambda)=\max \{\lambda_0e^{\beta \nu}, \lambda_{\max}\}$ and $\rho(y)=e^{-y^2/(2\sigma^2)}/\sqrt{2\pi \sigma^2}$. The physical running time on the supercomputer of Kyoto University was bounded at 20 hours. The number of parallel threads was 112 and the total simulation times were $T_{\mathrm{tot}}\approx 38777$ for $\eps^2=10^{-4}$, $T_{\mathrm{tot}}\approx 38830$ for $\eps^2=10^{-5}$, and $T_{\mathrm{tot}}\approx 23552$ for $\eps^2=10^{-6}$, by defining $T_{\mathrm{tot}}:=\sum_{i=1}^{112}T_i$ with simulation time on the $i$th thread. The other parameters are the same as those for Figure~\ref{fig:trajectory}.

\section{Technical note on calculations}\label{sec:detailed_calc}
	\subsection{On the Bachmann-Landau-like inequality notation}
	\label{sec:SM:Bachmann-Landau}
		In this Letter, the Bachmann-Landau equality and inequality notation is defined by 
		\begin{subequations}
		\begin{align}
			a(x) = O(b(x)) \>\>\> &\Longleftrightarrow \>\>\> \lim_{x \to \infty}\frac{a(x)}{b(x)} < \infty, \\
			a(x) > O(b(x)) \>\>\> &\Longleftrightarrow \>\>\> \lim_{x \to \infty}\frac{a(x)}{b(x)} = \infty, \\
			a(x) = o(b(x)) \>\>\> &\Longleftrightarrow \>\>\> \lim_{x \to \infty}\frac{a(x)}{b(x)} = 0.
		\end{align}
		Using this notation, we obtain 
		\begin{equation}
			a(x) > O(b(x)) \>\>\> \Longleftrightarrow \>\>\> 
			1/a(x) = o\left(1/b(x)\right)\>\>\> \Longleftrightarrow \>\>\> 
			b(x) = o(a(x))\label{eq:BL-O-inverting}.
		\end{equation}
		\end{subequations}

	\subsection{On the Laplace-like transformation}
		\label{sec:SM:LaplaceLikeTrans}
		In the main text, we have introduced a Laplace-like transformation
		\begin{equation}
			h(t) = \int_0^\infty dx \tilh(x) e^{-t/x}.
		\end{equation}
		This transformation is similar to the Laplace transformation. Indeed, by introducing the variable transformation $x:=1/s$, we obtain 
		\begin{equation}
			h(t) = \int_0^\infty ds H(s) e^{-ts} = \int_0^\infty dx \frac{1}{x^2}H\left(\frac{1}{x}\right) e^{-t/x}
		\end{equation}
		with the Laplace representation $H(s)$. This calculation implies that $\tilde{h}(x)$ is equivalent to $x^{-2}H(x^{-1})$. 

	\subsection{On the condition for explosive solutions}
		\label{sec:SM:existenceSol}
		Here we present an intuitive discussion on the existence of solutions for nonpositive mean mark $m\leq 0$. To capture intuitively the nature of the dynamics, let us truncate the Kramers-Moyal expansion~\eqref{SIeq:KM_negativeM} up to the second order:
		\begin{equation}
			\frac{\partial P_t(W;\bm{Z}')}{\partial t} \simeq \left[
				-m\frac{\partial }{\partial W}G(W;\bm{Z}')+
				\frac{\alpha_2}{2}\frac{\partial^2}{\partial W^2}G(W;\bm{Z}')
			\right]P_t(W;\bm{Z}), \>\>\> 
			\alpha_{2}:=\int_{-\infty}^\infty dy y^2\rho(y). 
		\end{equation}
		Recall that the function $G$ has been introduced in equation (\ref{SIeq:master_discrete}) as 
	$G(\bm{z}):= g\left(\sum_{k=1}^Kz_k\right)$.		
		This Fokker-Planck equation is equivalent to the following stochastic differential equation 
		\begin{equation}
			dW \simeq mG(W;\bm{Z}')dt + \sqrt{G(W;\bm{Z}')}dB_t, \>\>\> d\bm{Z}'=0 
			\label{SIeq:explosion_SDE}    
		\end{equation}
		with the standard Brownian motion $B_t$. For positive mean mark $m>0$, the time-evolution is explosive. Indeed, the noise term is negligible for large $W\to\infty$, 
		\begin{equation}
			|mG(W;\bm{Z}')| \gg \sqrt{G(W;\bm{Z}')}
		\end{equation}
		and the effectively deterministic dynamics
		\begin{equation}
			\frac{dW}{dt} \simeq mG(W;\bm{Z}')
		\end{equation} 
		is obviously explosive as soon as $G$ grows faster than linearly as a power law or exponential function. On the other hand, the negative-mean case $m<0$ is not explosive. 
		
		This intuitive discussion is consistent with the rigorous mathematical results on singular stochastic processes~\cite{ChernyB}. In addition, the rigorous results in Ref.~\cite{ChernyB} guarantee that the solution is not explosive even for the marginal case $m=0$. Indeed, according to \cite{ChernyB}, for an SDE 
		\begin{equation}
			dX_t = b(X_t)dt + \sigma(X_t)dB_t, \>\>\> X_0 = x_0,
		\end{equation}
		the classification of the solution is based on the following quantities
		\begin{equation}
			\rho_{\rm CE}(x) := \exp\left(-\int_a^x \frac{2b(y)}{\sigma^2(y)} dy\right), \>\>\>
			s_{\rm CE}(x) := -\int_x^\infty \rho(y)dy, \>\>\> 
			x \in [a,\infty)
		\end{equation}
		and 
		\begin{equation}
			\int_a^\infty \rho_{\rm CE}(x)dx, \>\>\> 
			\int_{a}^\infty \frac{|s_{\rm CE}(x)|dx}{\rho_{\rm CE}(x)\sigma^2(x)}
		\end{equation}
		with some appropriate $a$. For the SDE~\eqref{SIeq:explosion_SDE}, $\rho_{\rm CE}(x)$ and $s_{\rm CE}(x)$ are given by 
		\begin{align}
			\rho_{\rm CE} (x) &:= \exp\left(-\int_a^x \frac{2mG}{G} dy\right) = e^{-2m(x-a)}, \\ 
			s_{\rm CE}(x) &:= -\int_{x}^\infty dye^{-2m(y-a)} = 
			\begin{cases}
				-\infty & (m\leq 0) \\
				\frac{1}{2m}e^{-2m(x-a)} & (m > 0) \\
			\end{cases}.
		\end{align}
		According to \cite{ChernyB}, the non-positive-mean case with $m\leq 0$ is classified as {\it recurrent}, without explosion to $\infty$, because
		\begin{equation}
			\int_{a}^\infty \rho_{\rm CE}(y)dy =\infty \>\>\> (m\leq 0). 
		\end{equation}
		On the other hand, for positive mean mark $m>0$, the system is classified as {\it explosive}. Indeed, we obtain 
		\begin{equation}
			\int_{a}^\infty \rho_{\rm CE}(y)dy < \infty, \>\>\> \int_{a}^\infty \frac{|s_{\rm CE}(x)|dx}{\rho_{\rm CE}(x)G(x)} = \frac{1}{2m}\int_{a}^\infty \frac{dx}{G(x)} < \infty \>\>\> (m> 0),
		\end{equation}
		because $G^{-1}(x)$ decays faster than $x^{-2}$ on the assumption of the fast-accelerating intensity. 

	\subsection{Derivation of solution~\eqref{eq:sol_integral_eq_const}}\label{sec:Der_sol_integral_equation}
		Here we show that the solution of the integral equation
		\begin{equation}
			\int_{-\infty}^\infty dy\rho(y)\phi(\nu-y) - \phi(\nu) \simeq 0 \>\> \mbox{for large }\nu.
			\label{eq:integral_equation_appendix}
		\end{equation}
		Let us first assume that the solution is given by an exponenital $\phi(\nu) = e^{-c\nu}$. 
		By direct substitution, we obtain the self-consistent condition
		\begin{equation}
			\Phi(c) = 0, 
		\end{equation}
		by defining 
		\begin{equation}
			\Phi(x):= \int_{-\infty}^\infty dy\rho(y)(e^{xy}-1).
		\end{equation}
		Remarkably, $\Phi(x)$ is a strictly convex function because 
		\begin{equation}
			\frac{d^2\Phi(x)}{dx^2} = \int_{-\infty}^\infty y^2\rho(y)e^{xy}dy > 0. 
			\label{SIeq:CGF_convex}
		\end{equation}
		This means that $\Phi(x)$ has no more than one minimum.
		Here we assume that $\rho(y)$ decays sufficiently fast and $\Phi(x)$ exists.  
		The general solution of Eq.~\eqref{eq:integral_equation_appendix} is given by the superposition of exponentials (i.e., the two-sided Laplace representation),
		\begin{equation}
			\phi(\nu) \simeq \sum_{i} C_ie^{-c_i\nu}
		\end{equation}
		with the $i$th zero point $c_i$, satisfying $\Phi(c_i)=0$ and $c_i < c_{i+1}$. 

	\subsubsection{Assuming a zero-mean mark distribution}
		\begin{figure}
			\centering
			\includegraphics[width=150mm]{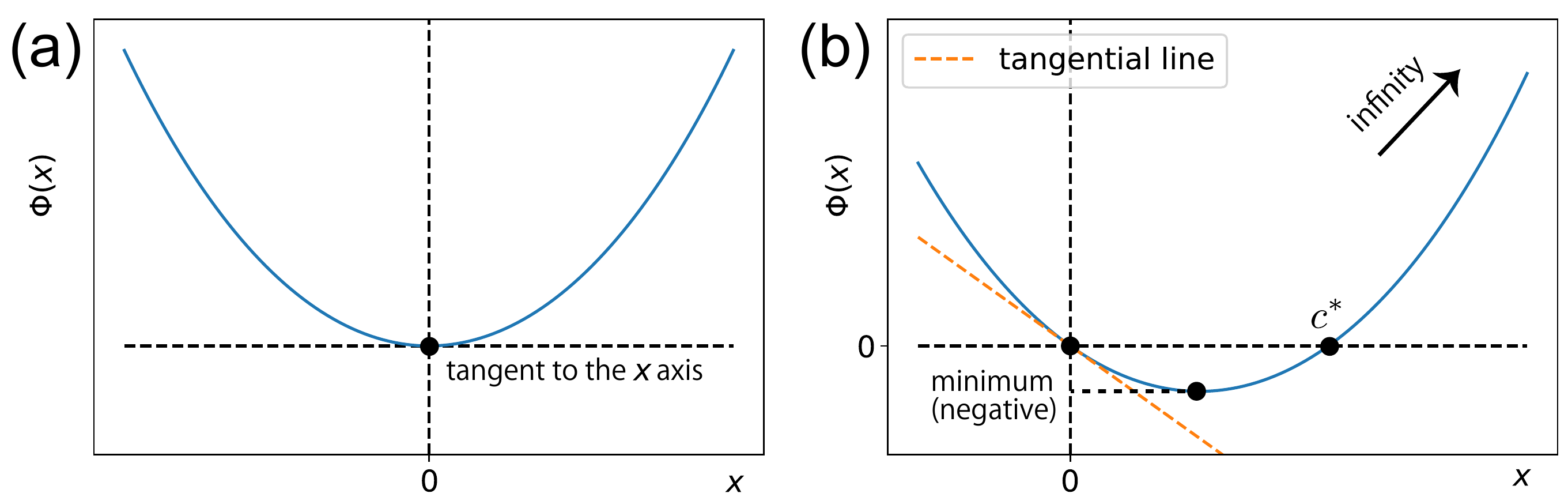}
			\caption{
								Schematic figure of the moment-generating function $\Phi(x)$. $\Phi(x)$ is a strictly convex function with specific values $\Phi(0)=0$, $\Phi(+\infty)=\infty$, and $d\Phi(0)/dt=m$. 
								(a)~Case with zero-mean mark $m=0$, where the curve is tangent to the $x$ axis at $x=0$. 
								(b)~Case with negative-mean mark $m<0$, where the tangential line at $x=0$ has negative coefficient $d\Phi(0)/dt=m<0$. Considering this geometrical shape, the minimum of $\Phi(x)$ occurs at a negative value of $\Phi(x)$ for $x>0$ and the roots of $\Phi(c)=0$ are given by $c=0$ and $c=c^*>0$. 
							}
				\label{SIfig:CGF}
		\end{figure}
		Let us assume that the mean mark is zero: 
		\begin{equation}
			m := \int_{-\infty}^\infty y\rho(y)dy = 0. 
		\end{equation}
		For the zero-mean mark distribution, the equation $\Phi(c)=0$ has a single root at $c=0$ (see Fig.~\ref{SIfig:CGF}a). Indeed, the minimum of $\Phi(x)$ is at $x=0$ because 
		\begin{equation}
			\frac{d\Phi(x)}{dx}\bigg|_{x=0} = \int_{-\infty}^\infty y\rho(y)dy = m = 0.
		\end{equation}
		Because the minimum value of $\Phi(x)$ is given by $\Phi(0)=0$, the only real solution of $\Phi(c)=0$ is therefore given by $c=0$. 
		
		Interestingly, $c=0$ is the double root of $\Phi(c)=0$ and thus a special treatment is necessary: One of the basic solutions of Eq.~\eqref{eq:integral_equation_appendix} is given by a constant function
		\begin{equation}
			\phi(\nu) \simeq \sum_{i} C_ie^{-c_i\nu} = C_0. 
		\end{equation}
		In addition, the affine function
		\begin{equation}
			\phi(\nu) \simeq C_0 + C_1 \nu 
			\label{SIeq:int_sol_gen_meanzero_solsol}
		\end{equation}
		with another constant $C_1$ is also a solution. Indeed, we obtain the consistent relation
		\begin{equation}
			\int_{-\infty}^\infty dy\rho(y)\left(C_0 + C_1 (\nu-y)\right) - \left(C_0 + C_1 \nu\right)
			= 0,
		\end{equation}
		by considering $\int_{-\infty}^\infty dy\rho(y)=1$ and $\int_{-\infty}^\infty y\rho(y)dy=0$. Notably, the affine form of the solution~\eqref{SIeq:int_sol_gen_meanzero_solsol} can be systematically derived by considering the zero-mean limit of the negative-mean case, as discussed below. 

	\subsubsection{Assuming a negative-mean mark distribution}
		Next, let us assume that the mean mark is negative and the probability of positive marks is nonzero
		\begin{equation}
			m := \int_{-\infty}^\infty y\rho(y)dy < 0, \>\>\>
			m_+ := \int_{0}^\infty y\rho(y)dy > 0. 
		\end{equation}
		Under this condition, the derivative of $\Phi(x)$ at $x=0$ is negative,
		\begin{equation}
			\frac{d\Phi(x)}{dx}\bigg|_{x=0} = \int_{-\infty}^\infty y\rho(y)dy = m < 0.
		\end{equation}
		In addition, $\lim_{x\to\infty} \Phi(x) = + \infty$ because 
		\begin{equation}
			\Phi(x) = \int_{0}^\infty dy\rho(y)(e^{xy}-1) + \int_{-\infty}^0dy\rho(y)(e^{xy}-1) 
			> \int_{0}^\infty dy\rho(y)xy + \int_{-\infty}^0dy\rho(y)(0-1) 
			= m_+x -p_- \to +\infty
		\end{equation}
		for $x\to +\infty$, where we have introduced $p_-:=\int_{-\infty}^0 \rho(y)dy$ and have used the following inequalities: $e^{xy}\geq xy + 1$ for $x\geq 0$ and $e^{xy}>0$ for any $x$. 
		
		Considering these properties, the schematic picture of $\Phi(x)$ is given by Fig.~\ref{SIfig:CGF}b and all the roots of $\Phi(c)=0$ are given by $c=0$ and $c=c^*>0$. We thus find that the solution of Eq.~\eqref{eq:integral_equation_appendix} is given by 
		\begin{equation}
			\phi(\nu) \simeq \sum_{i} C_ie^{-c_i\nu} = C_1 + C_0e^{-c^*\nu}. 
			\label{SIeq:int_sol_gen_negativeM_solsol}
		\end{equation}

		\paragraph*{For the zero-mean mark limit.}
			For a reference, let us consider the zero-mean mark limit $m\uparrow 0$. Interestingly, for infinitesimal negative $m$, the positive root of $\Phi(c)=0$ approaches zero, such that $c^*\downarrow 0$ for $m\uparrow 0$. Then, the solution~\eqref{SIeq:int_sol_gen_negativeM_solsol} can be expanded as 
			\begin{equation}
				\phi(\nu) \simeq \sum_{i} C_ie^{-c_i\nu} = C_1 + C_0 - c^*C_0\nu  + O(C_0c^{*2})
			\end{equation}
			up to the second order by assuming infinitesimal positive $c^*$. Here we replace $C'_0 := C_0+C_1$ and $C'_1:=-c^*C_0$ to obtain 
			\begin{equation}
				\phi(\nu) \simeq \sum_{i} C_ie^{-c_i\nu} = C'_0 + C'_1\nu  + O(C'_1c^{*1}).
			\end{equation}
			By taking the limit $m\uparrow 0$ and thus $c^*\downarrow 0$, we obtain the affine form of the solution~\eqref{SIeq:int_sol_gen_meanzero_solsol} 
			and the specific values of the constants $C'_0$ and $C'_1$.

	\subsubsection{Example: Gaussian mark distribution}
		\label{app:sec:Gaussian_mark}
		As an example, let us consider the case of the Gaussian mark distribution:
		\begin{equation}
			\rho(y) = \frac{1}{\sqrt{2\pi \sigma^2}}e^{-(y-m)^2/(2\sigma^2)}
		\end{equation}
		with mean $m$ and variance $\sigma^2$. The moment-generating function is given by 
		\begin{equation}
			\Phi(x) = \int_{-\infty}^\infty dy\rho(y)(e^{xy}-1) = e^{mx+\sigma^2x^2/2}-1.
		\end{equation}
		This means that the non-zero root of $\Phi(c^*)=0$ is given by 
		\begin{equation}
			c^* := -\frac{2m}{\sigma^2}. 
		\end{equation}
	
	\subsection{Derivation of Eq.~\eqref{SIeq:marginalization_calc}}
		\label{sec:app:Der_Marginalization}
		The integration in Eq.~\eqref{SIeq:marginalization_calc} can be performed as follows. From the definition~\eqref{SIdef:psi_from_phi} and the asymptotic solution~\eqref{eq:sol_integral_eq_const2}, the asymptotic steady state solution to the master equation~\eqref{SIeq:master_zeroMean} is given by 
		\begin{equation}
			P_{\mathrm{ss}}(\bm{z})=\left\{G(\bm{z})\right\}^{-1}\phi(\bm{z})=\left\{G(\bm{z})\right\}^{-1}\psi(W;\bm{Z}')\approx \left\{G(\bm{z})\right\}^{-1}C_0(\bm{Z}')
		\end{equation}
		with $G(\bm{z}):=g\left(\sum_{k=1}^K z_k\right)$, where the variable transformation $\bm{z}\to \bm{Z}:=(W,\bm{Z}')$ is defined by Eq.~\eqref{SIeq:var_trans_intEq}. We then obtain
		\begin{align}
			&\int_{-\infty}^{\infty} d\bm{z}P_{\mathrm{ss}}(\bm{z})\delta\left(\nu-\sum_{k=1}^{K}z_k\right) 
			\approx \int_{-\infty}^{\infty} d\bm{z}C_0(\bm{Z}')\left\{G(\bm{z}\right)\}^{-1}\delta\left(\nu-\sum_{k=1}^{K}z_k\right) \notag \\
			= &\int_{-\infty}^{\infty} d\bm{z}C_0(\bm{Z}')\left\{g\left(\sum_{k=1}^{K}z_k\right)\right\}^{-1}\delta\left(\nu-\sum_{k=1}^{K}z_k\right) 
			= \int_{-\infty}^{\infty} d\bm{z}C_0(\bm{Z}')\left\{g(\nu)\right\}^{-1}\delta\left(\nu-\sum_{k=1}^{K}z_k\right) \notag \\
			= & \left\{g(\nu)\right\}^{-1}   \int_{-\infty}^{\infty} dz_1\int_{-\infty}^{\infty} \left(\prod_{j=2}^{K}{dz_k}\right)C_0\left(z_2-\frac{\tilh_2}{\tilh_1}z_1,\dots ,z_K-\frac{\tilh_K}{\tilh_1}z_1\right)\delta\left(\nu-\sum_{k=1}^{K}z_k\right).
		\end{align}
		Here we apply a variable transformation $z_j':= z_j - \frac{\tilh_j}{\tilh_1}z_1$ for $j=2,\dots, K$ to obtain the relation
		\begin{align}
			&\int_{-\infty}^{\infty} dz_1\int_{-\infty}^{\infty} \left(\prod_{j=2}^{K}{dz_k}\right) C_0\left(z_2-\frac{\tilh_2}{\tilh_1}z_1,\dots, z_K-\frac{\tilh_K}{\tilh_1}z_1\right)\delta\left(\nu-\sum_{k=1}^{K}z_k\right) \notag \\
			=&\int_{-\infty}^{\infty} dz_1\int_{-\infty}^{\infty} \left(\prod_{j=2}^{K} dz_k'\right)C_0\left(z_2^\prime,\ldots,z_K^\prime\right)\delta\left(\nu-\sum_{k=2}^{K}z'_k-rz_1\right) \notag \\
			=&\int_{-\infty}^{\infty} \left(\prod_{j=2}^{K} dz'_k \right) C_0\left(z'_2, \dots, z'_K \right)\int_{-\infty}^{\infty} dz_1 \delta\left(\nu-\sum_{k=2}^{K}z'_k-rz_1\right).
		\end{align}
		Finally, by considering the identities for the $\delta$ functions
		\begin{equation}
			\delta(ax-b) = \frac{\delta(x-b/a)}{|a|}, \>\>\> \int_{-\infty}^\infty dz_1\delta(z_1-b) = 1
		\end{equation}
		for the constants $a\neq 0$ and $b$, we obtain 
		\begin{equation}
			\int_{-\infty}^{\infty} dz_1 \delta\left(\nu-\sum_{k=2}^{K}z'_k-rz_1\right) 
			= \frac{1}{r}\int_{-\infty}^{\infty} dz_1 \delta\left(z_1-\frac{\nu-\sum_{k=2}^{K}z'_k}{r}\right)
			=\frac{1}{r},
		\end{equation}
		which allows us to deduce
		\begin{equation}
			\int_{-\infty}^{\infty} d\bm{z}P_{\mathrm{ss}}(\bm{z})\delta\left(\nu-\sum_{k=1}^{K}z_k\right) \approx 
			{\left\{g(\nu)\right\}^{-1} \over r}\int_{-\infty}^{\infty} \left(\prod_{j=2}^{K} dz'_k \right) C_0\left(z'_2, \dots, z'_K\right)
		\end{equation}
		with a constant $r:= (1/\tilh_1)\sum_{k=1}^K \tilh_k$. This implies Eq.~\eqref{SIeq:marginalization_calc} by assuming $(1/r)\int_{-\infty}^{\infty} C_0(z'_2,\dots,z'_K)\Pi_{j=2}^Kdz'_j<\infty$.

	\subsection{Derivation of Eq.~\eqref{SIeq:marginalization_calc_negativeM}}
	\label{sec:app:Der_Marginalization_negativeM}
		The integration in Eq.~\eqref{SIeq:marginalization_calc_negativeM} can be performed as follows. From the definition~\eqref{SIdef:psi_from_phi} and the asymptotic solution~\eqref{SIeq:sol_master_negativeM2}, the steady state solution is given by 
		\begin{equation}
			P_{\mathrm{ss}}(\bm{z})=\left\{G(\bm{z})\right\}^{-1}\phi(\bm{z})=\left\{G(\bm{z})\right\}^{-1}\psi(W;\bm{Z}')=\left\{G(\bm{z})\right\}^{-1}C_0(\bm{Z}')e^{-c^*W}
		\end{equation}
		with $G(\bm{z}):=g\left(\sum_{k=1}^K z_k\right)$, where the variable transformation $\bm{z}\to \bm{Z}:=(W,\bm{Z}')$ is defined by Eq.~\eqref{SIeq:var_trans_intEq}. We then obtain
		\begin{align}
			&\int_{-\infty}^{\infty} d\bm{z}P_{\mathrm{ss}}(\bm{z})\delta\left(\nu-\sum_{k=1}^{K}z_k\right) 
			\approx \int_{-\infty}^{\infty} d\bm{z}C_0(\bm{Z}')e^{-c^*W}\left\{G(\bm{z}\right)\}^{-1}\delta\left(\nu-\sum_{k=1}^{K}z_k\right) \notag \\
			= &\int_{-\infty}^{\infty} d\bm{z}C_0(\bm{Z}')e^{-c^*W}\left\{g\left(\sum_{k=1}^{K}z_k\right)\right\}^{-1}\delta\left(\nu-\sum_{k=1}^{K}z_k\right) 
			= \int_{-\infty}^{\infty} d\bm{z}C_0(\bm{Z}')e^{-c^*W}\left\{g(\nu)\right\}^{-1}\delta\left(\nu-\sum_{k=1}^{K}z_k\right) \notag \\
			= & \left\{g(\nu)\right\}^{-1}   \int_{-\infty}^{\infty} dz_1\int_{-\infty}^{\infty} \left(\prod_{j=2}^{K}{dz_k}\right)C_0\left(z_2-\frac{\tilh_2}{\tilh_1}z_1,\dots ,z_K-\frac{\tilh_K}{\tilh_1}z_1\right)\exp\left(-\frac{c^*}{\tilh_1}z_1\right)\delta\left(\nu-\sum_{k=1}^{K}z_k\right).
		\end{align}
		Here we apply a variable transformation $z_j':= z_j - \frac{\tilh_j}{\tilh_1}z_1$ for $j=2,\dots, K$ to obtain the relation
		\begin{align}
			&\int_{-\infty}^{\infty} dz_1\int_{-\infty}^{\infty} \left(\prod_{j=2}^{K}{dz_k}\right) C_0\left(z_2-\frac{\tilh_2}{\tilh_1}z_1,\dots ,z_K-\frac{\tilh_K}{\tilh_1}z_1\right)\exp\left(-\frac{c^*}{\tilh_1}z_1\right)\delta\left(\nu-\sum_{k=1}^{K}z_k\right) \notag \\
			=&\int_{-\infty}^{\infty} dz_1\int_{-\infty}^{\infty} \left(\prod_{j=2}^{K} dz_k'\right)C_0\left(z_2^\prime,\ldots,z_K^\prime\right)\exp\left(-\frac{c^*}{\tilh_1}z_1\right)\delta\left(\nu-\sum_{k=2}^{K}z'_k-rz_1\right) \notag \\
			=&\int_{-\infty}^{\infty} \left(\prod_{j=2}^{K} dz'_k \right) C_0\left(z'_2, \dots, z'_K \right)\int_{-\infty}^{\infty} dz_1 \exp\left(-\frac{c^*}{\tilh_1}z_1\right)\delta\left(\nu-\sum_{k=2}^{K}z'_k-rz_1\right)
		\end{align}
		with a constant $r:= (1/\tilh_1)\sum_{k=1}^K \tilh_k$. Considering the relation for the $\delta$ function
		\begin{align}
			\int_{-\infty}^{\infty} dz_1 \exp\left(-\frac{c^*}{\tilh_1}z_1\right)\delta\left(\nu-\sum_{k=2}^{K}z'_k-rz_1\right) &= 
			\frac{1}{r}\int_{-\infty}^{\infty} dz_1 \exp\left(-\frac{c^*}{\tilh_1}z_1\right)\delta\left(z_1-\frac{\nu-\sum_{k=2}^{K}z'_k}{r}\right) \notag \\
			& = \frac{1}{r}\exp\left(-\frac{c^*}{\tilh_1}\frac{\nu-\sum_{k=2}^{K}z'_k}{r}\right),
		\end{align}
		we finally obtain 
		\begin{equation}
			\int_{-\infty}^{\infty} d\bm{z}P_{\mathrm{ss}}(\bm{z})\delta\left(\nu-\sum_{k=1}^{K}z_k\right) \propto 
			\left\{g(\nu)\right\}^{-1}\exp\left(-\frac{c^*}{\tilh_{\rm tot}}\nu\right)
		\end{equation}
		and $\tilh_{\rm tot} := r\tilh_1 = \sum_{k=1}^K \tilh_k$. This implies Eq.~\eqref{SIeq:marginalization_calc_negativeM} by assuming 
		\begin{equation}
			{1 \over r}\int_{-\infty}^{\infty} \left(\prod_{j=2}^{K} dz'_k \right) C_0\left(z'_2, \dots, z'_K\right)\exp\left(\frac{c^*}{\tilh_{\rm tot}}\sum_{k=2}^K z_k'\right)<\infty. 
		\end{equation}

\section{Author contributions}
	KK conceived the technical framework and performed the analytical and numerical calculations. DS designed the research, contributed to and checked the analytical calculations and supervised this project. KK and DS discussed all of the results, developed their interpretation and wrote the manuscript. 
\end{widetext}

\end{document}